\documentclass[twocolumn, twocolappendix]{openjournal}

\usepackage{xcolor}
\usepackage{textgreek}
\usepackage[utf8]{inputenc}
\usepackage[english]{babel}
\usepackage[T1]{fontenc}
\usepackage{afterpage}
\usepackage{amsmath}
\usepackage{amssymb}
\usepackage{booktabs}
\usepackage{caption}
\usepackage{color}
\usepackage{fdsymbol}
\usepackage{gensymb}
\usepackage{graphicx}
\usepackage{hyperref}
\usepackage{makecell}
\usepackage{multirow}
\usepackage{placeins}
\usepackage{rotating}
\usepackage{subfigure}
\usepackage{tabularx}
\usepackage{tikz}

\usepackage{hyperref}
\hypersetup{
    unicode, 
    colorlinks=true,
    linkcolor=linkcolor,
    citecolor=linkcolor,
    filecolor=linkcolor,
    urlcolor=linkcolor,
}
\usepackage{color,colortbl}
\definecolor{linkcolor}{rgb}{0.0,0.3,0.5}
\usepackage{tensind}
\tensordelimiter{?}
\DeclareGraphicsExtensions{.bmp,.png,.jpg,.pdf}
\usepackage{verbatim}
\newcommand{\diana}[1]{\textcolor{black}{#1}}

\usepackage{soul}

\graphicspath{{./figs/}}

\makeatletter
\@ifundefined{longtable*}{}{%
  \expandafter\let\csname longtable*\endcsname\relax
  \expandafter\let\csname endlongtable*\endcsname\relax
}
\makeatother

\begin{document}

\captionsetup[table]{singlelinecheck=false}
\usetikzlibrary{positioning, shapes.geometric, arrows}

\newcommand{\ds}{\ensuremath{\Delta\Sigma}}
\newcommand{\gt}{\ensuremath{\gamma_{\rm t}}}
\newcommand{\rp}{\ensuremath{r_{\rm p}}}
\newcommand{\mpch}{\ensuremath{h^{-1} \, \mathrm{Mpc}}}

\title[DESI-KiDS Photometric Redshift Calibration]{The Power of DESI for Photometric Redshift Calibration: A Case Study with KiDS-1000}
\author{D.~Blanco$^{1}$,
A.~Leauthaud$^{1}$\footnotemark[\dagger],
J.~U.~Lange$^{2}$,
A.~H.~Wright$^{3}$,
H.~Hildebrandt$^{3}$,
S.~Heydenreich$^{1}$,
D.~Ravulapalli$^{1}$,
J.~Ratajczak$^{4}$,
K.~S.~Dawson$^{4}$,
J.~McCullough$^{5}$,
B.~Dey$^{6}$,
J.~N.~Aguilar$^{7}$,
S.~Ahlen$^{8}$,
A.~Anand$^{7}$,
D.~Bianchi$^{9, 10}$,
C.~Blake$^{11}$,
D.~Brooks$^{12}$,
F.~J.~Castander$^{13,14}$,
T.~Claybaugh$^{7}$,
A.~Cuceu$^{7}$,
A.~de la Macorra$^{15}$,
J.~Della~Costa$^{16,17}$,
A.~Dey$^{17}$,
A.~Elliott$^{18}$,
N.~Emas$^{19}$,
S.~Ferraro$^{7,20}$,
A.~Font-Ribera$^{20}$,
J.~E.~Forero-Romero$^{22,23}$,
C.~Garcia-Quintero$^{24, 48}$,
E.~Gaztañaga$^{14,25}$,
S.~Gontcho~A~Gontcho$^{7,26}$,
G.~Gutierrez$^{27}$,
D.~Huterer$^{28}$,
M.~Ishak$^{29}$,
J.~Jimenez$^{21}$,
S.~Joudaki$^{30}$,
D.~Joyce$^{17}$,
S.~Juneau$^{17}$,
D.~Kirkby$^{31}$,
A.~Kremin$^{7}$,
A.~Krolewski$^{32,33}$,
C.~Lamman$^{18}$,
M.~Landriau$^{7}$,
L.~Le~Guillou$^{34}$,
M.~Manera$^{21,35}$,
A.~Meisner$^{17}$,
R.~Miquel$^{21,36}$,
J.~Moustakas$^{37}$,
S.~Nadathur$^{25}$,
J.~Newman$^{38}$,
W.~Percival$^{32,33,39}$,
A.~Porredon$^{30}$,
F.~Prada$^{40}$,
I.~P\'erez-R\`afols$^{41}$,
A.~Robertson$^{17}$,
G.~Rossi$^{42}$,
E.~Sanchez$^{30}$,
C.~Saulder$^{43}$,
A.~Semenaite$^{44}$,
D.~Schlegel$^{7}$,
H.~Seo$^{45}$,
J.~H.~Silber$^{7}$,
D.~Sprayberry$^{17}$,
G.~Tarl\'{e}$^{46}$,
B.~A.~Weaver$^{17}$,
R.~Zhou$^{7}$,
H.~Zou$^{47}$
\\
[6pt]
$^{1}$ Department of Astronomy and Astrophysics, University of California Santa Cruz, 1156 High Street, Santa Cruz, CA 95064, USA\\
$^{2}$ Department of Physics, American University, 4400 Massachusetts Avenue NW, Washington, DC 20016, USA\\
$^{3}$ Ruhr University Bochum, Faculty of Physics and Astronomy, Astronomical Institute (AIRUB), German Centre for Cosmological Lensing, 44780 Bochum, Germany\\
$^{4}$ Department of Physics and Astronomy, The University of Utah, 115 South 1400 East, Salt Lake City, UT 84112, USA\\
$^{5}$ Department of Astrophysical Sciences, Princeton University, Princeton, NJ 08544, USA\\
$^{6}$ Department of Astronomy \& Astrophysics, University of Toronto, Toronto, ON M5S 3H4, Canada\\
$^{7}$ Lawrence Berkeley National Laboratory, 1 Cyclotron Road, Berkeley, CA 94720, USA \\
$^{8}$ Department of Physics, Boston University, 590 Commonwealth Avenue, Boston, MA 02215 USA \\
$^{9}$ Dipartimento di Fisica ``Aldo Pontremoli'', Universit\`a degli Studi di Milano, Via Celoria 16, I-20133 Milano, Italy \\
$^{10}$ INAF-Osservatorio Astronomico di Brera, Via Brera 28, 20122 Milano, Italy \\
$^{11}$ Centre for Astrophysics \& Supercomputing, Swinburne University of Technology, P.O. Box 218, Hawthorn, VIC 3122, Australia
\\
$^{12}$ Department of Physics \& Astronomy, University College London, Gower Street, London, WC1E 6BT, UK
\\
$^{13}$ Institut d'Estudis Espacials de Catalunya (IEEC), c/ Esteve Terradas 1, Edifici RDIT, Campus PMT-UPC, 08860 Castelldefels, Spain
\\
$^{14}$ Institute of Space Sciences, ICE-CSIC, Campus UAB, Carrer de Can Magrans s/n, 08913 Bellaterra, Barcelona, Spain
\\
$^{15}$ Instituto de F\'{\i}sica, Universidad Nacional Aut\'{o}noma de M\'{e}xico,  Circuito de la Investigaci\'{o}n Cient\'{\i}fica, Ciudad Universitaria, Cd. de M\'{e}xico  C.~P.~04510,  M\'{e}xico
\\
$^{16}$ Department of Astronomy, San Diego State University, 5500 Campanile Drive, San Diego, CA 92182, USA
\\
$^{17}$ NSF NOIRLab, 950 N. Cherry Ave., Tucson, AZ 85719, USA
\\
$^{18}$ The Ohio State University, 191 West Woodruff Avenue, Columbus, OH 43210, USA
\\
$^{19}$ Centre for Astrophysics \& Supercomputing, Swinburne University of Technology, P.O. Box 218, Hawthorn, VIC 3122, Australia
\\
$^{20}$ University of California, Berkeley, 110 Sproul Hall \#5800 Berkeley, CA 94720, USA
\\
$^{21}$ Institut de F\'{i}sica d’Altes Energies (IFAE), The Barcelona Institute of Science and Technology, Edifici Cn, Campus UAB, 08193, Bellaterra (Barcelona), Spain
\\
$^{22}$ Departamento de F\'isica, Universidad de los Andes, Cra. 1 No. 18A-10, Edificio Ip, CP 111711, Bogot\'a, Colombia
\\
$^{23}$ Observatorio Astron\'omico, Universidad de los Andes, Cra. 1 No. 18A-10, Edificio H, CP 111711 Bogot\'a, Colombia
\\
$^{24}$ Center for Astrophysics $|$ Harvard \& Smithsonian, 60 Garden Street, Cambridge, MA 02138, USA
\\
$^{25}$ Institute of Cosmology and Gravitation, University of Portsmouth, Dennis Sciama Building, Portsmouth, PO1 3FX, UK
\\
$^{26}$ University of Virginia, Department of Astronomy, Charlottesville, VA 22904, USA
\\
$^{27}$ Fermi National Accelerator Laboratory, PO Box 500, Batavia, IL 60510, USA
\\
$^{28}$ Department of Physics, University of Michigan, 450 Church Street, Ann Arbor, MI 48109, USA
\\
$^{29}$ Department of Physics, The University of Texas at Dallas, 800 W. Campbell Rd., Richardson, TX 75080, USA
\\
$^{30}$ CIEMAT, Avenida Complutense 40, E-28040 Madrid, Spain
\\
$^{31}$ Department of Physics and Astronomy, University of California, Irvine, 92697, USA
\\
$^{32}$ Department of Physics and Astronomy, University of Waterloo, 200 University Ave W, Waterloo, ON N2L 3G1, Canada
\\
$^{33}$ Perimeter Institute for Theoretical Physics, 31 Caroline St. North, Waterloo, ON N2L 2Y5, Canada
\\
$^{34}$ Sorbonne Universit\'{e}, CNRS/IN2P3, Laboratoire de Physique Nucl\'{e}aire et de Hautes Energies (LPNHE), FR-75005 Paris, France
\\
$^{35}$ Departament de F\'{i}sica, Serra H\'{u}nter, Universitat Aut\`{o}noma de Barcelona, 08193 Bellaterra (Barcelona), Spain
\\
$^{36}$ Instituci\'{o} Catalana de Recerca i Estudis Avan\c{c}ats, Passeig de Llu\'{\i}s Companys, 23, 08010 Barcelona, Spain
\\
$^{37}$ Department of Physics and Astronomy, Siena University, 515 Loudon Road, Loudonville, NY 12211, USA
\\
$^{38}$ Department of Physics \& Astronomy and Pittsburgh Particle Physics, Astrophysics, and Cosmology Center (PITT PACC), University of Pittsburgh, 3941 O'Hara Street, Pittsburgh, PA 15260, USA
\\
$^{39}$ Waterloo Centre for Astrophysics, University of Waterloo, 200 University Ave W, Waterloo, ON N2L 3G1, Canada
\\
$^{40}$ Instituto de Astrof\'{i}sica de Andaluc\'{i}a (CSIC), Glorieta de la Astronom\'{i}a, s/n, E-18008 Granada, Spain
\\
$^{41}$ Departament de F\'isica, EEBE, Universitat Polit\`ecnica de Catalunya, c/Eduard Maristany 10, 08930 Barcelona, Spain
\\
$^{42}$ Department of Physics and Astronomy, Sejong University, 209 Neungdong-ro, Gwangjin-gu, Seoul 05006, Republic of Korea
\\
$^{43}$ Max Planck Institute for Extraterrestrial Physics, Gie\ss enbachstra\ss e 1, 85748 Garching, Germany
\\
$^{44}$ Centre for Astrophysics \& Supercomputing, Swinburne University of Technology, P.O. Box 218, Hawthorn, VIC 3122, Australia
\\
$^{45}$ Department of Physics \& Astronomy, Ohio University, 139 University Terrace, Athens, OH 45701, USA
\\
$^{46}$ University of Michigan, 500 S. State Street, Ann Arbor, MI 48109, USA
\\
$^{47}$ National Astronomical Observatories, Chinese Academy of Sciences, A20 Datun Road, Chaoyang District, Beijing, 100101, P.~R.~China
\\
$^{48}$ NASA Einstein Fellow
}

\footnotetext[\dagger]{E-mail: alexie@ucsc.edu}

\date{Accepted xxx. Received xxx}

\label{firstpage}

\begin{abstract}
    Accurate redshift estimates are a critical requirement for weak lensing surveys and one of the main uncertainties in constraints on dark energy and large-scale cosmic structure. In this paper, we study the potential to calibrate photometric redshift (photo-z) distributions for gravitational lensing using the Dark Energy Spectroscopic Instrument (DESI). Since its science operations began in 2021, DESI has collected more than 50 million redshifts, adding about one million monthly. In addition to its large-scale structure samples, DESI has also acquired over 256k high-quality spectroscopic redshifts (spec-zs) in the COSMOS and XMM/VVDS fields.  This is already a factor of 3 larger than previous spec-z calibration compilations in these two regions. 
    Here, we explore calibrating photo-z’s for the subset of KiDS-1000 galaxies that fall into joint self-organizing map (SOM) cells overlapping the DESI COSMOS footprint using the DESI COSMOS observations. Estimating the redshift distribution in KiDS-1000 with the new DESI data, we find broad consistency with previously published results, while also detecting differences in the mean redshift in some tomographic bins with an average shift of $\Delta\langle z \rangle = -0.028$ in the mean and $\Delta\tilde{z} = +0.011$ in the median between tomographic bins.
    However, we also find that incompleteness per SOM cell, i.e., groups of galaxies with similar colors and magnitudes, can modify the $n(z)$ distributions. Finally, we comment on the fact that larger photometric catalogs, aligned with the DESI COSMOS and DESI XMM/VVDS footprints, would be needed to fully exploit the DESI dataset and would extend the coverage to nearly eight times the area of existing 9-band photometry.
\end{abstract}

\maketitle

\section{Introduction}
The Dark Energy Spectroscopic Instrument (DESI; \citealt{DESICollaboration2016_arXiv_1611_0036}) is a next generation wide-field facility on the Mayall 4-meter telescope at Kitt Peak National Observatory, designed to measure the large-scale structure of the universe with high precision~\citep{DESICollaboration2016b, DESI_Collaboration_2022}. DESI employs a robotic fiber-positioning system that can simultaneously obtain nearly 5000 spectra across a $\sim 3^{\circ}$ field of view~\citep{Miller_2024, Poppett2024}, enabling an ambitious survey of roughly 17,000 deg$^2$ of sky over eight years~\citep{schlafly2024surveyoperationsdarkenergy}. The full DESI program is expected to yield approximately 63 million spectroscopically confirmed galaxies and quasars~\citep{Guy_2023}. The first Data Release (DR1;~\citealt{desicollaboration2025datarelease1dark}) includes spectra for more than 18 million unique targets and is already publicly available. Early DESI science results are already delivering competitive constraints on dark energy and cosmic acceleration~\citep{Abdul_Karim_2025, Adame_2025}. While its primary cosmological drivers are Baryon Acoustic Oscillations (BAO) and Redshift-Space Distortions (RSD), DESI’s unprecedented spectroscopic depth and diversity enable a much broader range of science, spanning galaxy evolution, quasar physics, the growth of structure, environmental studies, absorber systems, and a wide array of cross-survey calibration and legacy applications.
 
Gravitational lensing, the deflection of light from distant sources due to foreground mass distributions~\citep[e.g.,][]{Kilbinger2015, Dodelson2017, Mandelbaum_2018}, provides a powerful and complimentary observable. Weak lensing, in particular, has been identified as one of the most promising methods for probing the matter distribution in the Universe, offering strong constraints on parameters such as $\Omega_{\text{m}}$ and S$_8$, which are crucial for testing models of dark energy and structure~\citep{Albrecht2006reportdarkenergytask}. Because the distortions are small, a larger number of galaxies is necessary to detect the lensing effect. As a result, gravitational lensing surveys use photometric redshift (hereafter, ``photo-zs'') estimates based on broadband photometry instead of spectroscopic redshifts (hereafter, ``spec-zs''). Typical photometric redshift uncertainties for weak-lensing surveys are $\sigma_z / (1+z) \sim 0.02$–$0.05$ for well-measured samples, but even small mean biases of $|\Delta z| \sim 0.01$ can lead to several-percent errors in cosmological parameters~\citep[e.g.,][]{Hildebrandt2012_MNRAS_421_2355, Bonnett_2016, Mandelbaum_2018, Wright2020_AA_637_100}. Because photo-zs suffer from color-redshift degeneracies, they need to be calibrated~\citep[e.g.,][]{Newman_2022}. DESI can collect up to 5,000 spectra in one pointing, making it a true ``redshift machine''. The goal of this paper is to study DESI's potential for photo-z calibration for weak lensing science. A companion paper~\citep{lange2025} presents a new method for photo'z calibration using combinations of DESI redshifts and high fidelity photo-zs.

The Kilo-Degree Survey (KiDS, \citeauthor{DeJong2015} \citeyear{DeJong2015}) is a Stage-III gravitational lensing survey that has covered 1000 deg$^2$ in nine photometric bands. Cosmological parameter constraints from KiDS-1000 have been presented in~\cite{Asgari2021_AA_645_104} (cosmic shear), ~\cite{Heymans2021}~(3x2pt) and~\cite{Troster2022} (beyond $\Lambda$CDM) with the methodology presented in ~\cite{Joachimi2021}. To accurately understand gravitational lensing within KiDS and similar surveys, it is crucial to determine the intrinsic ensemble redshift distribution of source galaxies rather than relying on individual photometric estimates. However, these distributions can be biased and inaccurate when the underlying photometric redshifts suffer from incomplete or unrepresentative training sets, limited photometric coverage, or color–redshift degeneracies. Consequently, the inferred redshift distribution in such surveys depends on calibrating with smaller spectroscopic samples \citep{Hartley_2020}. Typically, this involves using a small, complementary subset of galaxies, which includes spec-z measurements from dedicated campaigns or more accurate photo-z estimates from deeper, higher-quality imaging surveys. For example, these surveys achieve photo-z uncertainties of $\sigma(z)/(1+z)\leq 0.006 - 0.010$ by combining high-S/N optical NIR imaging with intermediate- and narrow-band filters, which greatly reduces color-redshift degeneracies \citep[e.g.,][]{Laigle2016_ApJS_224_24, Weaver_2022}. 
One calibration approach used by KiDS \citep{Kuijken2019_AA_625_2, Hildebrandt2021_AA_647_124} utilizes self-organizing maps~\citep[SOMs,][]{kohonen1982}, building upon the methodology introduced by \citet{Masters2015} and \cite{Masters2017} and further developed in \citet{Buchs2019_MNRAS_489_820}, \cite{Myles2021_MNRAS_505_4249}, and \cite{Campos2024}. This technique has become a state-of-the-art standard in weak lensing analyses due to its effectiveness in mapping photometric color space and identifying spectroscopic gaps, making KiDS-1000 a widely used methodological reference point for SOM-based redshift calibration in weak-lensing analyses, though its representativeness is limited by KiDS’s shallower depth and broader wavelength coverage relative to surveys such as DES, HSC, and the forthcoming LSST. Ongoing efforts to optimize SOM calibration include targeted spectroscopic campaigns to fill underrepresented regions of color space \citep[e.g.][]{Masters2019_ApJ_877_81, Stanford2021_ApJS_256_9, McCullough2024}.
By employing the SOM technique, the redshift distribution is modeled as a function of the broad-band colors of source galaxies, effectively establishing a robust ``color-redshift'' relation. 
This established relationship is extended to the entire compilation of source galaxies, facilitating the redshift estimates for galaxies lacking direct spectroscopic measurements. 
The SOM technique additionally offers the capability of removing galaxies from the source sample if similar counterparts do not exist in the calibration samples, which reduces biases in the estimated redshift distribution. 
Approaches such as clustering redshifts, where the angular distribution of source galaxies is correlated with that of a reference sample with known redshifts \citep[e.g.,][]{Newman2008, Hildebrandt2021_AA_647_124}, and shear ratio methods \citep{Myles2021_MNRAS_505_4249, Sanchez2022} offer complementary pathways for estimating source redshift distributions.

Prior to the advent of DESI, photo-z calibrations have relied on spec-z compilations derived from a multitude of datasets. These datasets varied significantly in their observational strategies, instruments used, target selections, redshift success probability, and definitions of quality flags. As a result, these compilations often presented an inhomogeneous dataset, introducing challenges in maintaining consistent quality and reliability. The diversity in quality standards across different surveys and telescopes meant that calibrating photo-zs often involved navigating a complex landscape. This inhomogeneity poses significant challenges to ensuring that the spec-z compilations are not only accurate but also representative of the broader galaxy compilation~\citep{Myles2021_MNRAS_505_4249}. Consequently, as lensing strives to reach 1\% precision in key parameters such as S$_8$~\citep{LSSTDESC2021}, a representative and clean spectroscopic sample is crucial~\citep{Hildebrandt2021_AA_647_124}.

In this paper, we explore the potential of using DESI for photometric redshift calibration. Using KiDS as a case study, we revisit the source redshift distributions of the KiDS survey with new spectroscopic data from DESI. While the DESI Complete Calibration of the Color-Redshift Relation (DC3R2) marked the first use of DESI redshifts for redshift calibration, it was limited in depth and coverage \citep{McCullough2024}. In contrast, the sample used here is drawn from the deeper KiDS-COSMOS photometry and includes DESI ancillary observations that reach fainter magnitudes than the main survey. These data therefore have the potential to serve as an important and independent check on source redshift distributions, an essential input for robust dark energy constraints from weak lensing, though we do not directly perform cosmological inference in this work.

Our paper is structured as follows. Section~\ref{sec:data} describes the observational data used in our study. In Section~\ref{sec:methodology}, we describe the SOM methodology. Section~\ref{sec:results} presents our main results, including independent estimates of $n(z)$, an evaluation of how KiDS quality cuts and DESI completeness affect them, and a discussion of the implications. Finally, Section~\ref{sec:summary_and_conclusions} presents our summary and conclusions.

\section{Data}
\label{sec:data}

This Section outlines the observational data used for performing our photo-z calibration. The names of the catalogs used in this paper are presented in Appendix~\ref{app:A}.

\subsection{KiDS}
\label{subsec:KiDS} 

KiDS is a project specifically designed for detailed studies of weak gravitational lensing and galaxy formation. 
This work utilizes photometric observations from the fourth data release of the Kilo-Degree Survey \citep[KiDS DR4;][]{Kuijken2019_AA_625_2} which encompasses 1,006 square degrees of the sky divided into 1,006 individual survey tiles. Building on the foundation of the previous 440 tiles released in the KiDS DR3 catalog~\citep{deJong_2017}, the \textit{ugriZYJHK$_s$} images in this survey detect objects with limiting AB magnitudes corresponding to 5$\sigma$ detections in Gaussian Aperture and PSF (GAaP) photometry reaching up to 24.8 in the \textit{u}-band and 22.4 in the \textit{K$_s$}-band.

This work additionally incorporates the nine-band imaging dataset from the combined VISTA Kilo-degree INfrared Galaxy survey (VIKING) and KiDS programs, which when coupled with supplementary deep field observations, extends beyond the main survey depth. The KiDS observations deliver optical photometry in the \textit{ugri}~bands while VIKING contributes near-IR data in \textit{ZYJHK$_s$}. The combined footprint and wavelength coverage of the KiDS+VIKING instruments located at the ESO Paranal Observatory provide a broad wavelength range in the near-infrared (8,000 - 24,000 \AA) and optical wavelengths (3000 - 9,000~\AA). This expanded wavelength coverage, particularly in the near-infrared, improves photo-z estimates at higher redshifts. A detailed account of the combined KiDS dataset is offered in \cite{Wright2019_AA_632_34}. 

The KiDS-1000 dataset introduces a curated ``gold sample'' (hereafter referred to as S$_K$)
of galaxies, selected for their reliable shape and redshift measurements. Encompassing a total of 21,262,011 source galaxies, the gold sample is distinguished by its use of photo-z point estimates (z$_B$) derived through the Bayesian Photometric Redshift (BPZ) method \citep{Benitez2000_ApJ_536_571}, with a redshift probability prior informed by \cite{Raichoor2014}. These redshift estimates then define five tomographic bins, $Z_B \in (0.1, 0.3], (0.3, 0.5], (0.5, 0.7], (0.7, 0.9], (0.9, 1.2]$, as detailed in \cite{Kuijken2019_AA_625_2}. 

The redshift distribution of the gold sample is estimated using a set of overlapping spec-zs in combination with the self-organizing map methodology \citep{Wright2019_AA_632_34}. 
In this framework, the SOM is trained on the KiDS nine-band photometric color-magnitude space via an unsupervised dimensionality-reduction and clustering method. 
The SOM organizes galaxies into cells with similar photometric properties. An underlying assumption of SOMPZ is that spec-zs of galaxies that fall into a given cell provide an empirical estimate of the $p(z|c)$ relation of that cell.  Using the 25,373 spec-zs available in KiDS,  this training process enables the method to estimate the $n(z)$ within each SOM cell by analyzing the densities of the calibration and target samples in the photometric feature space. The SOM is also used to identify underrepresented or poorly calibrated regions by finding cells with no, or insufficient, spectroscopic coverage. In KiDS, `insufficient' spectroscopic coverage is defined using the QC1 quality criterion which identifies cells where the spec-z and photo-z relationship is unreliable. This implementation is described in Section~\ref{subsec:quality_cuts}.
Galaxies in these cells are excluded based on criteria that quantify deviations between the mean spec-z of the calibration sample and the mean photo-z of the target sample within each cell. Each tomographic bin is calibrated separately to ensure an unbiased estimation of the mean redshift of galaxies within those bins ~\citep{laureijs2011eucliddefinitionstudyreport}.  

The KiDS DR4 calibration uses the following spec-zs: an assortment of spectra from Chandra Deep Field South \citep[CDFS;][]{Popesso2019, Balestra2010}, DEEP2~\citep{Newman2013_ApJS_208_5}, the GAMA \citep[Galaxy And Mass Assembly,][]{Driver2011_MNRAS_413_971} deep field G15Deep \citep{Kafle2018_MNRAS_475_4434, Driver_2022}, VVDS~\citep[VIMOS VLT Deep Survey;][]{LeFevre_2005, LeFevre_2013}, and specz-s in COSMOS~\citep{Lilly2009_ApJS_184_218}. This collective data are summarized in Table~\ref{tab:kids_calibration_table}. 

The COSMOS dataset employed in this calibration includes a non-public catalog, provided by the COSMOS team. Although referred to by the KiDS team as ``zCOSMOS'', this dataset is, in reality, a comprehensive compilation of various catalogs within the COSMOS field \citep{Khostovan2025}. For clarity, we designate this as the ``COSMOS Team'' spec-zs in this paper. This compilation encompasses zCOSMOS \citep{Lilly_2007}, zCOSMOS Bright \citep[zCOSMOS-B;][]{Lilly2009_ApJS_184_218}, zCOSMOS Deep \citep[zCOSMOS-D;][]{Mignoli_2019}, among others, and we utilize the 2019 version of this dataset. In order to verify that our calibration matches that of KiDS, we first use the full KiDS spec-z calibration catalogs. Afterwards, for the comparison between the KiDS and DESI in COSMOS calibration, we only use the KiDS COSMOS Team spec-z catalog, which amounts to a total of 9,930 objects.

In the KiDS-1000 observations, the COSMOS field lacks VISTA z-band data. Consequently, to maintain consistent coverage across bands, the current COSMOS catalog combines KiDS \textit{ugri} photometry, VISTA \textit{YJHK$_s$} photometry that has been noise-degraded by the KiDS team to match the VIKING survey depth, and z-band photometry sourced from downgraded MegaCAM observations on the Canada-France-Hawaii Telescope \citep[CFHT;][]{Hildebrandt2009}. This uniform photometry is essential for assigning accurate color vectors to the KiDS calibration catalogues. In this work, all DESI and KiDS spectroscopic galaxies are assigned SOM colors using the deep KiDS-COSMOS 9-band photometry; because SOM calibration requires deep, homogeneous multi-band data, our DESI-KiDS comparison is therefore restricted to the inner $\sim$1 deg$^2$ COSMOS region where such photometry exists. Outside this footprint, DESI spectroscopy does not overlap with photometry of sufficient depth or band coverage to map DESI galaxies onto the KiDS SOM in a consistent way. For this reason, all DESI-KiDS comparisons in this work are performed only within the region of shared deep photometric support.   

 \begin{table*}
    \centering
    \begin{tabular}{llclr
    rrr}
    \hline \hline
    Type & Sample & Area & Selection & No. in & No. in & Total \\
         &        & [deg$^2$] && COSMOS & XMM/VVDS   & \\ 
    [0.5ex]
    \hline
    \multirow{5}{*}{KiDS Spectroscopy ($S_{K}$)} 
         & CDFS   & 0.1       & CDFS \& ESO-GOODS & 0      & 0              & 2~044 \\
         & DEEP2        & 0.8       & \texttt{ZQUALITY $\geq$ 3 \&} $z_{error} <$ 1\% & 0      & 0              & 6~919 \\         
         & GAMA-G15Deep & 1.0       &\texttt{Z\_QUAL $\geq$ 3 \&} $z > 0.001 $ & 0      & 0              & 1~792 \\
         & VVDS         & 1.0       & \makecell[l]{WIDE, DEEP, UDEEP \\ 
                                    \texttt{\& ZFLAGS $\in$ \{3,4,23,24\}}} & 0 & 4~688 &  4~688 \\ 
        & COSMOS Team & $\sim$0.5 &  & 9~930 &  0  & 9~930 &  \\    
    \hline
    \textbf{All KiDS} & & & & & & \textbf{25~373} \\
    \hline
    \end{tabular}
    \caption{Samples of spec-zs employed by the KiDS survey to calibrate redshifts summarized in \protect\cite{Wright2019_AA_632_34}. Note: zCOSMOS, zCOSMOS-B, and zCOSMOS-D are part of the COSMOS Team compilation \citep{Khostovan2025}.}
    \label{tab:kids_calibration_table}
\end{table*}

\begin{table}
    \centering
    \begin{tabular}{ll}
    \hline
    \hline
    Type & Selection \\
    \hline
    BGS & \texttt{SPECTYPE == `GALAXY'}\\ 
    & \texttt{\&} \texttt{z < 1.5} \\
    & \texttt{\&} \texttt{SNR\_z > 1.1} \\
    & \texttt{\&} \texttt{CHI2 $\geq$ 7000} \\
    & \texttt{\&} \texttt{DELTACHI2 $\geq$ 100 OR 40 $\leq$ DELTACHI2 $<$ 100 } \\
        \hline
    LRG & \texttt{SPECTYPE == `GALAXY'}\\ 
    & \texttt{\&} \texttt{z < 1.6} \\
    & \texttt{\&} \texttt{SNR\_z > 0.7} \\
    & \texttt{\&} \texttt{CHI2 $\geq$ 7000} \\
    & \texttt{\&} \texttt{DELTACHI2 $\geq$ 100 OR 40 $\leq$ DELTACHI2 $<$ 100 } \\
        \hline
    ELG & \texttt{SPECTYPE == `GALAXY'}\\ 
    & \texttt{\&} \texttt{z < 1.6} \\
    & \texttt{\&} \texttt{SNR\_r > 0.23} \\
    & \texttt{\&} \texttt{DELTACHI2 $\geq$ 25} \\
    & \texttt{\&} \texttt{log10(OII\_SNR) > 0.9 - 0.2log10(DELTACHI2)} \\
        \hline
    \end{tabular}
    \caption{Recommended quality cuts for BGS, LRG, and ELG target types as described in \citet{Ratajczak2025}.}
    \label{tab:desi_quality_cuts}
\end{table}

\begin{table*}
    \small
    \centering
    \begin{tabularx}{\textwidth}{llllrrr}
    \hline \hline
    Type        & Sample        & Reference                     & Selection                                     & No. in & No. in & Total \\
                &               &                               &                                                   & COSMOS & VVDS   &       \\
    [0.5ex]
    \hline
    \multirow{15}{*}{\makecell[l]{\\ Non-DESI \\ Spectroscopy \\ ($S_{ND}$)}}  
                & eBOSS ELG     &\cite{Raichoor2021_MNRAS_500_3254}                               & \texttt{zwarn $=$ 0 \& spectype $\ne$ `STAR'}     & 0      & 167     & 167  \\
                
                & eBOSS LRG     &\cite{Ross2020_MNRAS_498_2354} & \makecell[l]{\texttt{zwarning $=$ 0} \\ \texttt{\& class\_redrock $\ne$ `STAR'}}  
                                                                                                                    & 0      & 30      & 30   \\
                
                & SDSS DR16     & \cite{Ahumada2020_ApJS_249_3} & \texttt{zwarning $=$ 0 \& class $\ne$ `STAR'}     & 530    & 974     & 1~504\\
                
                & OzDES         & \citet{Lidman2020_MNRAS_496_19}& \texttt{2 $\leq$ qop $<$ 6}                      &  0     & 1~059   & 1~059\\
                
                & MMT/Binospec  & \cite{Fabricant2019_PASP_131_5004}& \texttt{redshift\_final $\ne$ NaN}            & 93     & 0       & 93   \\
    
                & C3R2 DR3      &\cite{Stanford2021_ApJS_256_9} & \texttt{Qual $\geq$ 3}                            & 333    & 304     & 637  \\
    
                & VIPERS        & \cite{Scodeggio2018_AA_609_84} & \texttt{2 $\leq~$zflg $<$ 10}                    & 0      & 1~876   & 1~876\\
    
                & GAMA DR3      & \cite{Baldry2018_MNRAS_474_3875}& \texttt{nQ $>$ 2}                               & 0      & 1~147   & 1~147\\
                
                & DEIMOS        & \cite{Hasinger2018_ApJ_858_77}& \texttt{Qf $\geq$ 3}                              & 5~849  & 0       & 5~849\\
    
                & LEGA-C DR2    & \cite{Straatman2018_ApJS_239_27}& \makecell[l]{\texttt{Fint $=$ 0 \& f\_z $=$ 0 \& FSp~$=$ 0} 
                                                                                        \\ \texttt{\& Use $=$ 1}}   & 1~354  & 0       & 1~354\\
    
                & MOSDEF        & \cite{Kriek2015_ApJS_218_15}  & \makecell[l]{\texttt{target $=$ 1 \& z\_mosfire $>$ 0} 
                                                                                 \\ \texttt{\& z\_mosfire\_zqual}}  & 236    & 0       & 236 \\

                & VVDS          & \cite{LeFevre2015_AA_576_79}  & \texttt{zflags $\notin\,\lbrace 0, 20 \rbrace$}  & 0      & 954     & 954 \\
                
                & DEEP2         & \cite{Newman2013_ApJS_208_5}  & \texttt{badflag $=$ 0 \& zquality $\geq$ 3 }.     & 0      & 8~889   & 8~889\\
                
                & zCOSMOS DR3   & \cite{Lilly2009_ApJS_184_218} & \texttt{zflag $>$ 2}                              & 8~413  & 0       & 8~413\\
    \textbf{All Non-DESI} &      &                               &                        & \textbf{16~808} & \textbf{15~400} & \textbf{84~355}\\
    \hline
                & COSMOS Team\textsuperscript{$\dagger$}   & \makecell[l]{\citet{Khostovan2025}} &                                                & 97~929  & 0     & 97~929\\    
    \hline
    \textbf{All DESI} ($S_{D}$)&& \makecell[l]{\citet{Ratajczak2025} 
                                                                    \\} 
                                                                & \makecell[l]{\texttt{quality\_z = 1, dz < 0.0033,}
                                                                \\
                                                                \texttt{quality\_z = 1, dz \textless{} 0.0033,}
                                                                \\ \texttt{`MWS', or `QSO'}} 
                                                                                         & \textbf{190~099} & \textbf{66~658} & \textbf{256~757} \\
    [0.5ex]
    \hline
    \end{tabularx}
    \begin{flushleft}
\footnotesize
\textsuperscript{$\dagger$}\,\diana{Listed for context and later comparison only. This catalog includes DESI EDR spectra and therefore partially overlaps with the DESI sample.} 
\end{flushleft}
    \caption{\diana{Calibration catalogs referenced in this work.} The existing Non-DESI high-quality redshifts correspond to targets with C3R2 photometry. DESI spec-zs are a compilation of redshifts within DESI tiles that overlap with COSMOS and XMM/VVDS, using a combination of fiducial DESI quality cuts and additional signal-to-noise requirements to reduce the number of catastrophic failures, as detailed in \citet{Ratajczak2025}. This table includes only objects that satisfy \texttt{quality\_z = 1}, and the full set of cuts is provided in Table~\ref{tab:desi_quality_cuts}. In this work, we have also removed stars, including Milky Way Survey targets, and QSOs. The COSMOS Team entry\textsuperscript{$\dagger$} is included for context but ingests the DESI EDR and is therefore not part of the Non-DESI baseline.}
    \label{tab:calibration_table}
    
\end{table*}

\subsection{Non-DESI Spectroscopy in COSMOS and XMM/VVDS}
\label{subsec:non-desi}

We compare the number of DESI spectra to a non–DESI compilation of redshifts in COSMOS and XMM/VVDS. Here, ``XMM” refers to the XMM–LSS field (the XMM–Newton Large-Scale Structure survey region; RA $\approx 35^\circ$, Dec $\approx -5^\circ$) with extensive multi-wavelength and spectroscopic follow-up \citep{Pierre2004, Read2004, Chiappetti2005}, and “VVDS” refers to the VIMOS VLT Deep Survey fields that overlap this region, which provide large, magnitude-selected spectroscopic samples \citep{LeFevre_2005, LeFevre_2013}. 

Our non–DESI compilation, explicitly excluding any DESI observations,  contains 16,808 galaxies in the COSMOS field and 15,400 in XMM/VVDS, including observations by the C3R2 Survey ~\citep[Complete Calibration of the Color-Redshift Relation;][]{Masters2017, Masters2019_ApJ_877_81, Stanford2021_ApJS_256_9}. Additionally, the updated COSMOS Team compilation \citep{Khostovan2025} reports 97,929 unique spectroscopic redshifts in COSMOS but ingests the DESI EDR and is therefore listed separately and not included in our non–DESI baseline to avoid double counting. We refer to the non–DESI compilation as $S_{ND}$ hereafter.

\diana{We note that no explicit object-level deduplication is performed between $S_{\rm ND}$ and the DESI sample $S_{\rm D}$. A small overlap is therefore possible because DESI may have re-observed galaxies previously targeted by SDSS/eBOSS or other predecessor surveys. The purpose of $S_{\rm ND}$ in this work is to provide a descriptive comparison between DESI and the spectroscopy that was previously available in COSMOS and XMM/VVDS, rather than to construct a mutually exclusive master spectroscopic catalog. Any such overlap would therefore only affect the catalog-count comparison in Table~\ref{tab:calibration_table}, and would slightly overestimate the size of the pre-DESI spectroscopic baseline. It does not affect the main SOM-based calibration analysis, which compares $S_{\rm D}$ to the KiDS COSMOS spectroscopic calibration sample in shared SOM cells.}

\subsection{DESI}
\label{subsec:desi}

The DESI focal plane system is mounted on the four-meter Mayall telescope on Kitt Peak, Arizona. It is equipped with 5,000 robotically actuated fibers capable of simultaneously capturing spectra at a wavelength range of 3600 - 9800 \AA~\citep{DESICollaboration2016_arXiv_1611_0036,  DESICollaboration2016b}.   

This work uses all the DESI data collected up to April 2024 from the COSMOS and XMM/VVDS fields (hereafter referred to as S$_D$). These observations include the main DESI program targets but also a large number of  objects targeted through various ancillary efforts. In particular, the COSMOS and XMM/VVDS regions have a large number of additional spectra targeted via DESI ancillary programs. 
Data were selected to lie within the following RA and Dec cuts, expressed in degrees: XMM/VVDS at \( 33.5 < \text{RA} < 37.5 \) and \( -7 < \text{Dec} < -3 \), and COSMOS at \( 148 < \text{RA} < 152 \) and \( 0 < \text{Dec} < 4 \). These RA and Dec boundaries partially overlap with the KiDS COSMOS footprint, which covers $\sim$1 deg$^2$ in the COSMOS region; in contrast, the DESI spectra analyzed in this paper span an area approximately 16 times larger.

\diana{For COSMOS and XMM/VVDS, we adopt the field-specific redshift-quality criteria of \citet{Ratajczak2025}, summarized in Table~\ref{tab:desi_quality_cuts}, rather than the global DESI fiducial thresholds. These adopted cuts were designed specifically for the COSMOS and XMM/VVDS fields and were tuned using repeat-observation consistency and visual inspection. In DESI, redshift confidence metrics are largely based on the $\Delta\chi^2$ statistic, defined as the difference between the best-fit $\chi^2$ value and that of the second-best solution returned by the redshift-fitting pipeline. Larger $\Delta\chi^2$ values indicate a more secure and well-constrained redshift solution \citep{Guy_2023, Myers_2023}. The Ratajczak et al. criteria combine $\Delta\chi^2$ with additional per-band signal-to-noise requirements for continuum-dominated samples, and use a sliding [O~II] S/N--$\Delta\chi^2$ relation together with an (r)-band S/N floor for ELGs. For comparison, typical global DESI fiducial criteria include cuts such as \texttt{ZWARN = 0}, z<1.5, and $\Delta\chi^2>15$ for LRGs, or \texttt{ZWARN = 0} and $\Delta\chi^2>40$ for BGS. Relative to such fiducial criteria, the field-specific cuts adopted here reduce the implied catastrophic redshift-failure rate at nearly unchanged completeness, from 1.4\% to 0.72\% for BGS and from 0.56\% to 0.29\% for LRGs, while yielding $\gtrsim99.3\%$ cross-program consistency in these fields. This higher-purity selection is essential for photo-(z) calibration, where catastrophic spectroscopic failures could bias the inferred n(z) within SOM cells.}

In this analysis, we additionally excluded stars, including Milky Way Survey targets and quasars (QSOs).  Table~\ref{tab:desi_quality_cuts}
provides the recommended quality cuts for different object types, such as galaxies, from the main program. For concision in Table~\ref{tab:calibration_table}, these quality cuts are summarized by the flag \texttt{quality\_z}. For quality cuts related to objects from ancillary programs, please refer to \citet{Ratajczak2025}.
 
After removing stars and QSOs and imposing the quality metrics described in \citet{Ratajczak2025}, S$_D$ contains a total of 190,099 redshifts in COSMOS and 66,658 in XMM/VVDS. 

In total, the DESI catalog in COSMOS and XMM/VVDS comprises 256,757 high-quality galaxy spectra, a considerable increase from S$_K$, and S$_{ND}$. However, the areas covered by DESI and KiDS are different, as shown in Figure~\ref{fig:kids_desi_cosmos_footprint}. This will be discussed in more detail in Section~\ref{subsec:desi_and_kids_spec-z_comparison}. When limiting the DESI COSMOS catalog to the KiDS COSMOS footprint of \(\sim1 \text{ deg}^2\), DESI contains 19,866 spec-zs, and KiDS COSMOS contains 9,930 as shown in Figure~\ref{fig:kids_desi_cosmos_footprint}. The DESI COSMOS RA and Dec definitions extend beyond the KiDS COSMOS footprint, such that 170,233 of the DESI COSMOS spec-zs lie outside of this area. The non-DESI redshifts and DESI redshifts in the COSMOS and XMM/VVDS fields, including their origins, quality cuts, and quantities, are summarized and compared in Table~\ref{tab:calibration_table}.

\subsection{Advantages and Challenges of the DESI Redshift Measurements}
\label{sec:advantages_and_challenges}

 Unlike preceding spectroscopic calibration catalogs that often compile data from various telescopes and surveys with inconsistent quality assessments, the DESI catalog benefits from homogeneous and clearly defined quality flags. This is exemplified by our adoption of \citet{Ratajczak2025} selections for high-confidence redshifts, which involve customized quality cuts for each target class. These cuts are based on spectral properties such as signal-to-noise ratio (SNR), continuum strength, emission line strength from \texttt{FastSpecFit}~\citep{Moustakas2023}, emission line diversity, and goodness-of-fit parameters from \texttt{Redrock}, a tool developed by Bailey et al. (in prep). The adoption of this conservative metric is aimed at minimizing outliers and thus enhancing the reliability of the data sample. In this work, we define a spectroscopic observation as ``successful'' if it passes the quality cuts outlined in Table~\ref{tab:desi_quality_cuts}.  

Moreover, published catalogs from previous efforts tend to only include objects with successful redshifts, making it challenging to quantify the impact of redshift failures. In our work, we explore how redshift failures inform the reliability of our dataset and contribute to systematic differences in redshift distributions. One potential factor influencing these differences is the DESI redshift success rate at higher redshifts. The DESI design is optimized for detecting the [O II] doublet (3726.032, 3728.815~\AA). However, beyond \( z \approx 1.6 \), this feature shifts out of the DESI spectral range due to the 9700~\AA~filter edge, limiting the redshift success rate. This does not imply that extending DESI's wavelength range would make its galaxy population resemble KiDS, but rather that limited spec-z completeness at high z constrains how well DESI can calibrate the high-z tail of the KiDS $n(z)$. 

A dictionary of all samples mentioned here is summarized in Table~\ref{tab:data_defs}.

\begin{table}
    \centering
    \begin{tabular}{ll}
    \hline
    \hline
    Sample & Definition \\
    \hline
    $S_{\rm K}$ 
      & The KiDS spec-zs used in the original \\
      &  KiDS SOM calibration of \citet{Wright2020_AA_637_100} \\
    \hline
    $S_{\rm D}$ 
      & The DESI spec-zs in the COSMOS and XMM/ \\
      &  VVDS fields of~\citet{Ratajczak2025} \\
    \hline
    $S_{\rm ND}$ 
      & Previously obtained spec-zs from external \\
      &  surveys for targets with C3R2 photometry
    \\ 
    \hline
    \end{tabular}
    \caption{
    Summary of the spectroscopic samples used throughout this work.}
    \label{tab:data_defs}
\end{table}

\section{Methodology}
\label{sec:methodology}

In this section, we describe our methodology for calibrating photometric redshifts using self-organizing maps (SOMs) and applying them to KiDS and DESI data.

 \subsection{Redshift Calibration Using Self-Organizing Maps}
 \label{subsec:SOM_approach}

In this work, we constrain the intrinsic redshift distribution of the source population, $n(z)$, as a function of uncalibrated photometric estimates $\hat z_s$, using a ``calibration sample”: a subset of galaxies with both photometric estimates $\hat z_s$ and high-quality spectroscopic redshifts $z_s$. We distinguish object-level from population-level quantities. For an individual galaxy, $z_s$ denotes its secure spectroscopic (``truth”) redshift and $\hat z_s$ its photometric estimate. Using the calibration sample, we infer $n(z)$ for the full source sample (e.g., within each tomographic bin or $\hat z_s$ interval) and summarize it by the population mean $\langle z\rangle$ and median $\tilde z$. Unless stated otherwise, $\langle z\rangle$ and $\tilde z$ refer to the ``source-population” $n(z)$, not to the calibration subset and not to per-galaxy values.

In principle, this estimation is straightforward. However, using the calibration sample without adjustment can introduce biases, as high-quality redshift measurements are often preferentially obtained for specific galaxy populations—such as those that are particularly bright, have strong emission or absorption features, or occupy certain regions of the complex galaxy color-space landscape~\citep{Myles2021_MNRAS_505_4249}.
Thus, our calibration sample must be appropriately weighted to correct for these selection effects~\citep{Lima2008_MNRAS_390_118, Gruen2017, Hartley_2020}.

Several approaches have been proposed in the literature to perform this weighting procedure via a direct calibration~\citep[e.g.,][]{Lima2008_MNRAS_390_118, Hildebrandt2017_MNRAS_465_1454, Buchs2019_MNRAS_489_820, Wright2020_AA_637_100}. This work will use SOMs to estimate suitable weights for the calibration sample. We work within the existing KiDS-1000 SOM framework, assigning both DESI and KiDS spectroscopic galaxies to its pre-defined cells using the deep KiDS-COSMOS photometry.

A SOM is an unsupervised machine-learning algorithm that transforms higher-dimensional data into a lower-dimensional, in our case two-dimensional, manifold~\citep{kohonen1982}. In particular, pairs of objects with small separation in the SOM also have a small separation in the high-dimensional space. This allows us to define cells in the SOM of galaxies with similar properties. Ultimately, weights for the calibration sample are chosen so that the SOM (weighted) occupation of the calibration sample is representative of the full source sample used for gravitational lensing studies.

We follow recent approaches in the literature and use multi-band colors and magnitudes as the input data for SOMs~\citep{Wright2020_AA_637_100, VanDenBusch2022}. Colors are tightly correlated with redshift, and galaxies with similar colors tend to occupy similar redshift ranges. However, especially in shallower photometry with larger measurement noise, redshift can still show residual trends with apparent magnitude within a fixed-color SOM cell~\citep{Masters2019_ApJ_877_81, Speagle2019_MNRAS_490_5658}. \citet{McCullough2024} demonstrated this explicitly: the magnitude–redshift dependence at fixed color is small when using low-noise, deep COSMOS-UltraVISTA photometry, but becomes substantial in the noisier KiDS-VIKING wide-field data. In higher-noise regimes, measured color is a less precise estimator of intrinsic color, which allows magnitude-dependent selection effects to re-emerge even after color-based clustering~\citep{McCullough2024}. Thus, weighting the calibration sample by its distribution in a shallow galaxy color SOM might still lead to significant biases. To address this,~\cite{Wright2020_AA_637_100} and \cite{vandenBusch2020_AA_642_200} demonstrated that implementing additional cleaning and selection procedures improves calibration accuracy by reducing the bias caused by photometric noise. This approach involves creating a gold sample, which includes only those galaxies in the photometric sample that are adequately represented in the calibration sample. Here, adequate representation means that the SOM cell containing a given photometric galaxy has sufficient spectroscopic coverage to provide a reliable redshift calibration. Cells lacking representation are excluded after categorizing a photometric sample into SOM cells. Additional refinement is achieved by applying hierarchical clustering to the SOM, grouping cells with similar photometric properties instead of treating individual cells independently. This approach allows for a more nuanced trade-off, reducing the number of photometric sources excluded as a result of the partitioning of the high-dimensional color space while mitigating the bias introduced by the misrepresentation of the gold sample. In this work, we follow the same KiDS procedure and apply the KiDS gold-sample selection for all SOM-based calibration and comparison steps.

Our photo-z calibration samples can be categorized into two types: a ``wide'' selection with possibly shallow photometry that encompasses source galaxies used for gravitational lensing studies
and a ``calibration'' sample of high-quality spec-zs. We utilize the photometry and SOM color vectors courtesy of the KiDS team and reconstruct the KiDS SOM using the broad-band photometric colors of all galaxies, enabling us to discretize the colors of each galaxy analogous to the construction of the gold sample. Here, the broad-band colors refer to the nine KiDS+VIKING bands (ugriZYJHK$_s$) and their adjacent-band color differences, matching the KiDS-1000 SOM definition.

Finally, we note that the redshift distribution estimation and calibration procedure within KiDS includes a (post-measurement) shift of the redshift distributions, applied through the application of informative priors on the bias of each tomographic bin's $\langle{\rm z}\rangle$. These informative priors are derived for each tomographic bin using high-fidelity simulations of the KiDS wide and calibration samples, and are highly sensitive to the distribution of both wide and calibration sources used, and so must be recomputed when either the calibration sample or the sample of wide sources is modified \citep[see e.g.,][]{Wright2020_AA_637_100}. In KiDS-1000 these priors introduce maximal shifts in the $n(z)$ at the level of $|\delta \langle z\rangle| \leq 0.037$ at $95\%$ confidence \citep{Hildebrandt2021_AA_647_124}. In this work we opt not to apply these bias estimates to the KiDS-1000 redshift distributions, nor to recompute these informative priors for each of our calibration samples and wide-sample subsets. Instead we opt to investigate the direct impact of estimating $n(z)$ distributions for (common subsets of) KiDS-1000 wide sources with various calibration samples, agnostic to the simulation-inferred biases that these $n(z)$ would require in a full cosmological analysis. As such, we caution the reader against speculation over the possible changes to cosmological parameter estimates that one might find for KiDS-1000 when using our estimated $n(z)$, compared to the fiducial $n(z)$ presented by KiDS-1000.

Throughout this work, our goal is to estimate the redshift distribution $n(z)$ of the photometric sample by learning the conditional distribution $p(z|c)$, where $c$ denotes a galaxy's position in color-magnitude space. The spectroscopic calibration sample of KiDS or DESI provides an empirical estimate of this conditional relation, which we denote as $p(z_K|c)$ and $p(z_D|c)$, respectively. Differences between these two estimates reflect differences in the spectroscopic selection functions $p(c|z_K)$ and $p(c|z_D)$ rather than differences in the underlying photometric sample itself. Making this distinction explicit clarifies how comparing the KiDS and DESI calibration samples informs the SOMPZ redshift calibration procedure.

\subsection{Application to KiDS and DESI}

The wide weight \(w_{{\rm w},i}\) for each galaxy \(i\) in the calibration sample ensures that the weighed calibration sample is representative of the wide sample, i.e., it corrects for biases in the redshift selection. This weight is a function of the galaxy's wide color-magnitude \(c_i\) and is given by
\begin{equation}
    w_{{\rm w},i} \;=\; \frac{N_{\rm w}(c_i)}{N_{\rm z}(c_i)} \, ,
\end{equation}
where \(N_{\rm w}\) and \(N_{\rm z}\) denote the abundance of galaxies with \(c_i\) in the wide and calibration samples, respectively.  Here, \(N(c_i)\) is shorthand for the count in the SOM cell that contains \(c_i\). Equivalently, if \(j_i\) is the SOM cell index for galaxy \(i\), then \(w_{{\rm w},i} = N_{\rm w}(j_i)/N_{\rm z}(j_i)\).

\diana{The transfer weight $w_{{\rm t},i}(j)$ describes the probability that a galaxy $i$ in the calibration sample enters the wide sample and is assigned to wide SOM cell $j$, given its observed wide color-magnitude vector $c_i$ and the wide selection $\hat{s}$:
\begin{equation}
w_{{\rm t},i}(j) = p\left(j \mid \hat{s}, c_i\right).
\end{equation}
In the general SOMPZ formalism, this term corresponds to the transfer function between the calibration/deep photometry and the wide photometric sample, analogous to $p(c_k \mid s,c_i)$ in \citet{Myles2021_MNRAS_505_4249}. In this work, because the KiDS deep fields overlap the wide layer, we evaluate this probability deterministically from the observed wide photometry. Each calibration galaxy that has valid wide photometry and passes the wide selection cut $\hat{s}$ is assigned to a single wide SOM cell, $j_i$. We therefore implement
\begin{equation}
w_{{\rm t},i}(j) =
\mathbb{I}\left(\hat{s}*i=1\right),
\mathbb{I}\left(j=j_i\right),
\end{equation}
where $\mathbb{I}$ is the indicator function. Thus, $w*{{\rm t},i}(j)=1$ for the observed wide SOM cell of a calibration galaxy that passes the wide selection, and $w_{{\rm t},i}(j)=0$ otherwise.}

\diana{A more general treatment could estimate $p(j \mid \hat{s},c_i)$ as a stochastic transfer matrix by marginalizing over photometric noise, blending, and selection effects. Previous work has used simulated source injection or Monte Carlo resampling to estimate this relationship more robustly and mitigate shot noise from limited overlap samples \citep[e.g.,][]{Myles2021_MNRAS_505_4249, Everett2022}. We do not perform such an injection-based transfer-function calibration here. Instead, our deterministic implementation should be interpreted as the empirical transfer implied by the directly observed KiDS wide-layer photometry.}

Both \(w_{{\rm w},i}\) and \(w_{{\rm t},i}(j)\) are calculated for each galaxy \(i\) in the calibration sample. Specifically, \(w_{{\rm w},i}\) depends on the wide-band color of galaxy \(i\) through its SOM cell \(j_i\), while \(w_{{\rm t},i}(j)\) represents the probability of a galaxy being associated with a specific SOM cell \(j\). Combining these weights, each galaxy in the calibration sample is assigned a weight \(w_{{\rm SOM},i}(j)\), where \(j\) refers to the SOM cell index. This weight is given by
\begin{equation}
    w_{{\rm SOM},i}(j) \;=\; w_{{\rm w},i}\, w_{{\rm t},i}(j).
\end{equation}

\subsection{KiDS SOM-Based Quality Cuts}
\label{subsec:quality_cuts}

In our analysis, we adopt the quality control checks (`QC') applied by the KiDS team to minimize the bias introduced by photometric noise in the calibration process.

The approach of the KiDS team involves two primary sets of quality control checks. The first set, referred to as ‘QC1’, flags and removes spectroscopic-to-photometric groupings, i.e., collections of SOM cells evaluated together, that are catastrophic outliers in the distribution of photo-zs versus SOMz's (i.e., the mean redshift of the SOM grouping), using the criterion  
\begin{equation}
\frac{\left|\left\langle z_{\text{spec}} \right\rangle - \left\langle Z_B \right\rangle \right|}{\operatorname{nMAD}\left(\left\langle z_{\text{spec}} \right\rangle - \left\langle Z_B \right\rangle \right)} > 5.
\label{eq:QC1}
\end{equation}
Here, $Z_B$ is the KiDS Bayesian photometric redshift estimate derived from BPZ, and the means $\left\langle z_{\text{spec}} \right\rangle$ and $\left\langle Z_B \right\rangle$ are computed over each grouping of SOM cells being assessed. 
The nMAD is the normalized median absolute deviation, defined as  
\begin{equation}
\operatorname{nMAD}(x) = 1.4826 \times \text{median}(|x - \text{median}(x)|),
\end{equation}
scaled such that it approximates the standard deviation in the case of Gaussian noise. Here, \( \sigma_{\text{mad}} \) is defined as \( \sigma_{\text{mad}} = \operatorname{nMAD}(\langle z_{\text{spec}} \rangle - \langle Z_B \rangle) \), representing the spread of the residuals between the calibration sample redshift and the photo-z. The denominator of Equation~\ref{eq:QC1} is exactly this quantity, so Equation~\ref{eq:QC1} can be written equivalently as $\left| \langle z_{\text{spec}} \rangle - \langle Z_B \rangle \right| > 5\,\sigma_{\text{mad}}$.

\diana{This threshold of $>5$ is adopted directly from the KiDS SOM redshift-calibration analysis and is not re-tuned in this work. In the original KiDS implementation, this criterion was chosen to reject groupings of SOM cells where the mean calibration-sample redshift, $\langle z_{\rm spec}\rangle$, and the mean KiDS photo-$z$, $\langle Z_B\rangle$, differ significantly, indicating catastrophic disagreements. \citet{Wright2020_AA_637_100} showed that applying this criterion reduces the SOM redshift bias while removing only a small fraction of the source sample. The corresponding scatter,
$$
\sigma_{\rm mad} =
{\rm nMAD}\left(\langle z_{\rm spec}\rangle - \langle Z_B\rangle\right),
$$
was calibrated in the KiDS analysis using the fiducial spectroscopic calibration sample, yielding $\sigma_{\rm mad}\approx0.12$. This same value was found by the KiDS team to be appropriate for their validation cases and across the five tomographic bins used in the KiDS analysis.}

\diana{In this work, we apply the KiDS QC1 criterion, including the threshold $\sigma_{\rm mad}\approx0.12$, directly to the DESI-based calibration sample without modification. We do not re-evaluate or optimize this threshold for DESI. This choice is intentional: our goal is to compare DESI and KiDS calibration samples within the same SOM-selection framework, rather than to define a new DESI-specific QC criterion.}

This quality cut is applied after running BPz (Bayesian Photometric Redshift) to assign initial photo-zs for galaxies, which are subsequently grouped into SOM cells based on their photometric properties. These groupings are then refined using hierarchical clustering, which clusters SOM cells with similar photometric properties rather than filtering individual cells. This technique allows for a more nuanced trade-off between the number of photometric sources discarded due to the segmentation of the high-dimensional color space and the bias introduced by the misrepresentation of the gold sample.

\subsection{Reproducing the KiDS Results}

\begin{figure*}
\centering
\includegraphics[width=\textwidth]{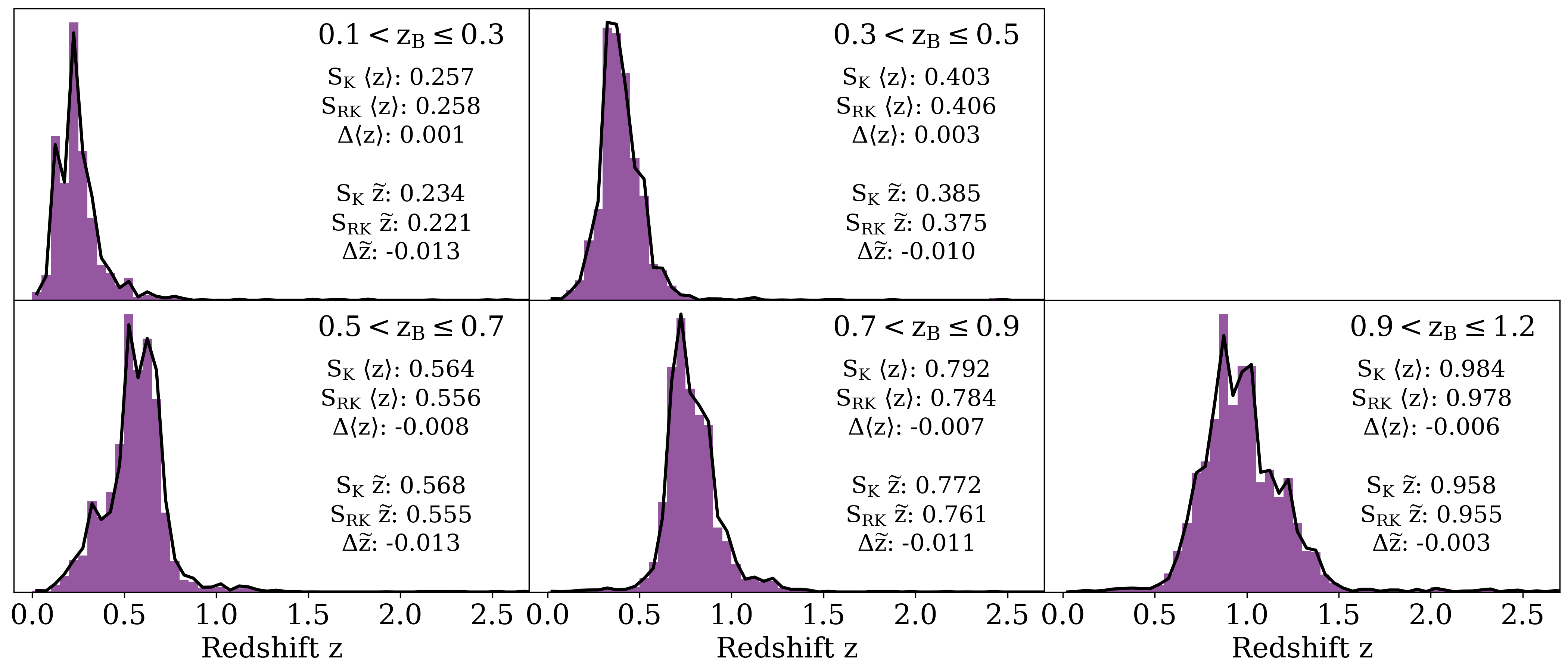}
\caption{Results of a sanity test to ensure that our SOM code can reproduce the KiDS results. We display the published redshift distributions ($n(z)$) from the KiDS survey in black. Overlayed in purple are our recreated $n(z)$s, reconstructed using our SOM methodology and the original KiDS calibration data (no DESI data). Each panel corresponds to a specific tomographic bin, composed of a subset of galaxies within the specified redshift range between 0.1 $\leq Z_B \leq$ 1.2. The values for the mean ($\langle z \rangle$) and median ($\tilde{z}$) redshifts are listed for both the published KiDS data (S$_K$) and our reconstruction (S$_{RK}$).}
\label{fig:mean_med_KiDS_recreation}
\end{figure*}

As a sanity check, we have replicated the KiDS published results following the methodology outlined by~\cite{Wright2020_AA_637_100}. This replication involved applying the SOM approach detailed in Section~\ref{subsec:SOM_approach} and using the KiDS spec-z compilation referenced in Table~\ref{tab:kids_calibration_table}. Our analysis covered the five tomographic bins ranging from $z=0.1$ to $z=1.2$.  

Figure~\ref{fig:mean_med_KiDS_recreation} compares the published KiDS $n(z)$ (black) to our SOM-based replication using the full KiDS calibration catalog (purple). 
Although we observe small bin-level differences in the mean redshift, we do not assess these differences using the full KiDS prior width as a significance test, because the KiDS priors and our replication are highly correlated. Instead we report the fractional shift relative to the KiDS prior width. The largest difference corresponds to only 0.3-0.4$\sigma$ of the KiDS redshift-bias prior, indicating that small methodological differences between our implementation and the KiDS pipeline are unlikely to meaningfully affect downstream weak-lensing analyses.  

The redshift-bias priors used for both KV450 and KiDS-1000 have Gaussian widths of $\sigma\simeq 0.010$--$0.012$ per bin (constructed from end-to-end simulations; \citealt{Wright2020_AA_637_100}), and were conservatively inflated (by doubling the simulation $\sigma$) in the cosmology analysis. 
These priors describe the simulation-based uncertainty relevant for cosmological inference, not the effective uncertainty when comparing two correlated estimates of n(z). 
The two-sided 97.5\% quantile of these priors is therefore $\approx 1.96\,\sigma \simeq 0.02$--$0.025$. These priors reflect uncertainties from photometric noise, spectroscopic selection/incompleteness, sample variance, and Poisson sampling in the calibration fields, rather than being a direct “measurement error” on the published KiDS $n(z)$ \citep[][App.~B]{Wright2020_AA_637_100}. For this replication, we therefore focus on the fractional shift described above, rather than applying the full prior width as a comparison threshold. Within this context, our differences are consistent with expectations. 

\diana{We emphasize that the uncertainty budget associated with a SOM-based redshift calibration is not captured by Poisson noise in the reconstructed n(z) alone. In the KiDS analyses, the uncertainty on the mean redshift of each tomographic bin is estimated using end-to-end simulations of the wide and calibration samples, which account for photometric noise, sample variance, Poisson sampling, spectroscopic selection effects, and redshift or magnitude incompleteness \citep[][App.~B]{Wright2020_AA_637_100}. These simulations produce informative priors on the residual redshift bias. Consequently, a formal uncertainty estimate for the DESI-based n(z)s presented in this work would require recomputing the same type of simulation-based calibration for the DESI selection function, redshift-success properties, and the specific wide-sample subsets considered here.}

\diana{We therefore do not interpret the shifts reported in this paper as cosmology-ready redshift-bias parameters. Instead, ($\Delta\langle z\rangle$) and ($\Delta\tilde{z}$) are used as compact diagnostics of the difference between redshift distributions inferred from the KiDS and DESI calibration samples in the same SOM cells. This distinction is important because the KiDS-1000 redshift calibration procedure includes an additional simulation-based recalibration of the measured n(z)s, while the DESI-based comparisons shown here are raw SOM estimates. A dedicated simulation campaign could in principle reduce, reweigh, or otherwise reinterpret some of the differences reported here, especially if they arise from differences in photometric-noise properties, sample selection, or redshift incompleteness between the calibration samples. Such a recalibration is beyond the scope of this pilot study, but is required before DESI-based n(z)s can be used directly in weak-lensing cosmological inference.}

As an additional cross-check, 99.98\% of KiDS spec-$z$ objects are assigned to their original KiDS SOM cells in our pipeline (a 0.02\% reassignment rate), indicating that our SOM implementation and association logic reproduce the published mapping. 
The small residual bin-level differences can arise from benign numerical effects, including floating-point precision, library-specific implementations of the distance metric, slight differences in cell-edge handling and association logic, and finite-sample fluctuations within individual cells. All of these lie well within the simulation-derived prior widths used by KiDS.

\section{Results}
\label{sec:results}

In this Section, we present our main results. We focus on SOM cells that are populated both by the DESI and KiDS spectra, and investigate 
$n(z)$'s derived from both datasets.

\diana{In later comparisons where the fiducial KiDS calibration sample is used as the baseline, the reference sample is instead $S_{K,\rm fid}$. We therefore interpret the sign of $\Delta\langle z\rangle$ consistently as ``comparison minus reference,'' with the reference sample specified in the relevant text, table, or figure caption.}

As discussed in Section 3.1, these comparisons do not include the simulation-based SOM bias calibration applied in the KiDS analyses; our results should therefore be interpreted as the raw, uncorrected n(z) estimates, without the post-measurement shifts used in the KiDS cosmological pipelines.

\subsection{Comparing the DESI and non-DESI data}
\label{subsec:desi_vs_non-desi}
\begin{figure}
\centering
\includegraphics[width=\columnwidth, trim=0cm 0cm 0cm 0cm]{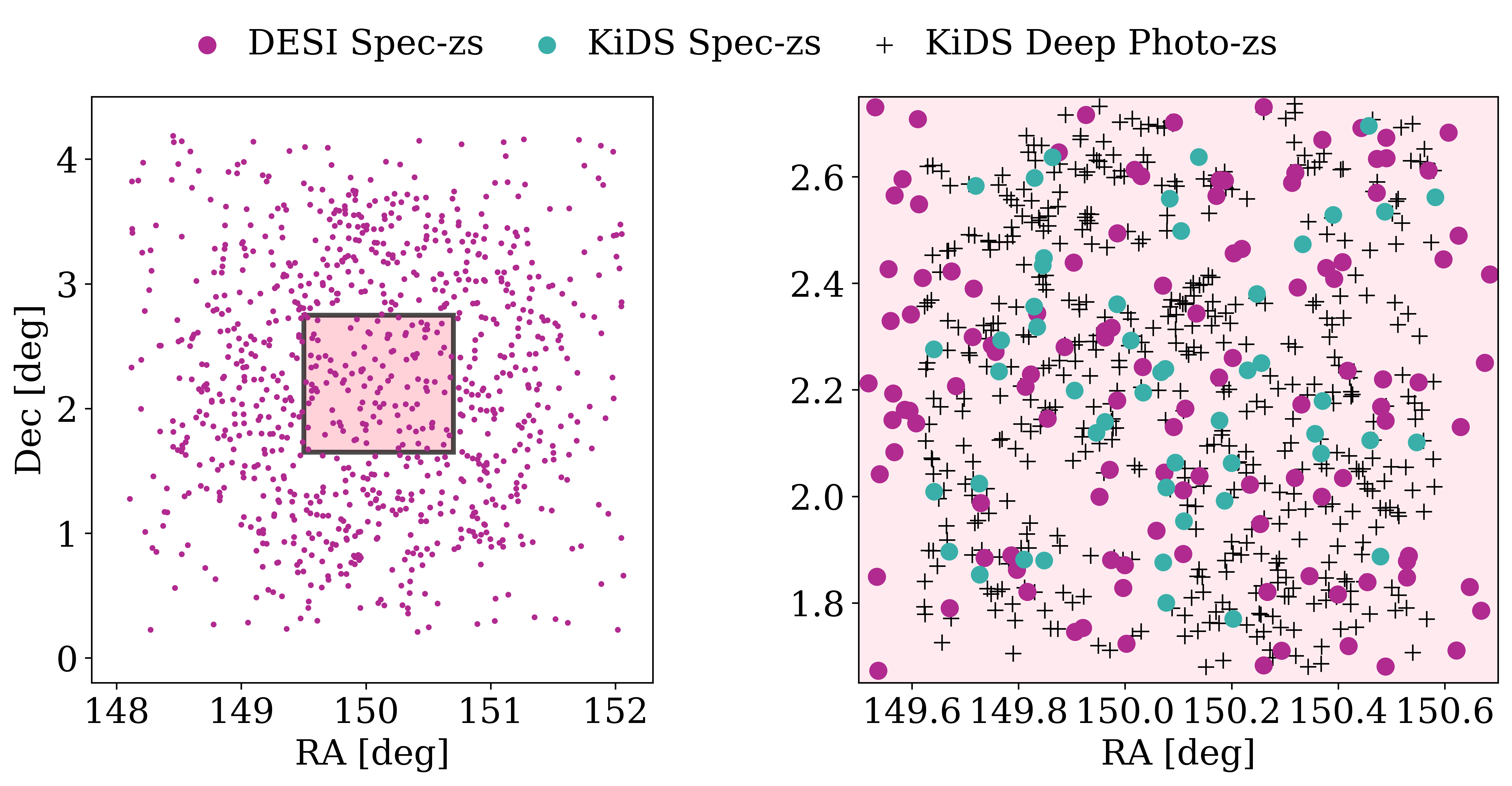}
\caption{Spatial distribution of the DESI COSMOS spectra (purple), the KiDS COSMOS spectra (teal), and the KiDS deep photometry in the COSMOS field (black crosses).
A random 0.5\% subsample of each catalog is shown, yielding 117 DESI spec-zs, 50 KiDS spec-zs, and 502 KiDS deep photo-z galaxies within the shared region highlighted by the pink square, which is enlarged in the right-hand panel. The KiDS COSMOS samples occupy only a fraction of the DESI footprint in the COSMOS field. The KiDS-1000 deep photometric catalog does not overlap with the XMM/VVDS field and is therefore not shown.
Within the inner shared area, DESI has a total of 19,866 spec-z’s, whereas KiDS has a total 9,930 spec-z’s.}
\label{fig:kids_desi_cosmos_footprint}
\end{figure}

Table~\ref{tab:calibration_table} indicates that DESI has obtained over 278,000 spec-zs in the COSMOS and XMM/VVDS fields combined. The size of the DESI dataset in the COSMOS field exceeds that of the KiDS COSMOS spec-z catalog by a factor of 19 and surpasses the COSMOS Team compilation of over 52,000 spec-zs by a factor of 5.

Figure~\ref{fig:kids_desi_cosmos_footprint} illustrates the spatial overlap between the DESI and KiDS spec-zs, and the KiDS-1000 deep photometry, shown through a random sampling of 0.5\% of all galaxies. Despite the extensive spectroscopic sampling achieved by DESI in the COSMOS region, its utility for photo-z calibration is constrained by the limited overlap with the KiDS COSMOS shared area and as well as the DESI depth. Specifically, the DESI catalog contains 19,866 redshifts within the KiDS COSMOS footprint, compared to 9,930 redshifts in the KiDS spec-z catalog. Thus, the DESI sample in this shared region is approximately 2 times larger than the KiDS sample.

The lack of deep, unified, and multi-band photometric data outside the KiDS COSMOS footprint limits the potential for leveraging DESI's extensive dataset. The SOM methodology employed for photo-z calibration relies on the availability of deep photometric data to distinguish galaxy colors and accurately establish the color-redshift relationship, establish the color–redshift relationship. Consequently, the disparity in coverage and depth between DESI and KiDS-1000, as shown in Figure~\ref{fig:kids_desi_cosmos_footprint}, represents a challenge in fully exploiting the DESI dataset for calibration purposes. 

To maximize the full potential of DESI in the future it will be necessary to incorporate deep photometry across these regions. The Hyper Suprime-Cam (HSC) wide and deep fields emerge as an alternative for a deep photometric sample. KiDS-1000 operates in the nine-band photometric system of \textit{ugriZYJHK$_s$}, which is broader than the \textit{grizy} five-band system of HSC, which could lead to a loss of information critical to analyses dependent on the more general wavelength coverage of KiDS. Future work will need to focus on strategies to supplement the DESI regions with deep photometry that matches the quality of KiDS. In particular, obtaining more uniform and extensive deep optical and near-infrared photometry across the XMM/VVDS and COSMOS fields will greatly enhance the utility of DESI for redshift calibration. Surveys such as LSST, Euclid, and Roman are expected to significantly expand coverage in these regions. Notably, one of the Euclid Deep Field South overlaps with the XMM-LSS/SXDS region and will provide deep public data in its first release, while Roman will offer complementary NIR coverage in regions overlapping with LSST. In addition, Roman's High-Latitude Wide-Area Survey is being designed to maximize overlap with major ground-based spectroscopic programs, including DESI, to enable joint cosmological analyses~\citep{Zasowski2025}.

\subsection{Comparison between the KiDS and DESI spectroscopic calibration catalogs}
\label{subsec:desi_and_kids_spec-z_comparison}

We compare the distributions of the KiDS and DESI spectroscopic calibration catalogs in SOM space. We focus hereafter on regions in SOM space that have spectroscopic redshift from both catalogs. Specifically, Joint S\(_K\) denotes the subset of KiDS spec-\(z\) galaxies that lie in SOM cells containing at least one KiDS and at least one DESI spec-\(z\), while Joint S\(_D\) analogously denotes the DESI galaxies in those same shared cells. In these joint SOM cells, KiDS and DESI provide independent empirical estimates of the color-redshift relation $p(z|c$). Differences between these samples therefor reflect differences in their spectroscopic selection functions $p(c|z)$ rather than differences in their underlying photometric population. Comparing Joint S$_K$ and Joint S$_D$ thus highlights where the two surveys sample the same regions of color-magnitude space in similar, or systematically different, ways.
By restricting to these joint cells, we also ensure a direct comparison between the two surveys in the same regions of SOM space, avoiding biases introduced by non-overlapping areas. We present redshift distributions in regions of common coverage. Our goal is to perform a pilot study and to quantify these joint distributions rather than perform cosmological inference. For this, please refer to our companion paper~\citep{lange2025}. These two catalogs are further described in Section~\ref{sec:data}.

Mapping the DESI targets onto the $ugriZYJHK_s$ color space provides a statistical sample instrumental for constraining the color-redshift relationship, which is vital for upcoming cosmological studies. Initially, the dataset comprised 190,099 DESI spec-z in COSMOS. However, calibration and subsequent analysis necessitate that the DESI targets possess consistent KiDS COSMOS Deep photometry, which is crucial for assigning accurate color-magnitude vectors. As described in Subsection \ref{subsec:desi} and~\ref{subsec:desi_vs_non-desi}, the DESI footprint is larger than the KiDS COSMOS field. Restricting the DESI catalog to the KiDS COSMOS footprint reduces the DESI COSMOS dataset to 19,866 objects. 

Restricting the analysis to the shared SOM cells between KiDS COSMOS and DESI COSMOS further reduces the number of available DESI redshifts, as we only consider SOM cells jointly occupied by both datasets. This final selection results in 8,557 DESI redshifts spanning 2,748 shared SOM cells or 26.9\% map coverage.

\begin{figure}
\centering
\includegraphics[width=\columnwidth, trim=0cm 0.8cm 0cm 0.5cm]{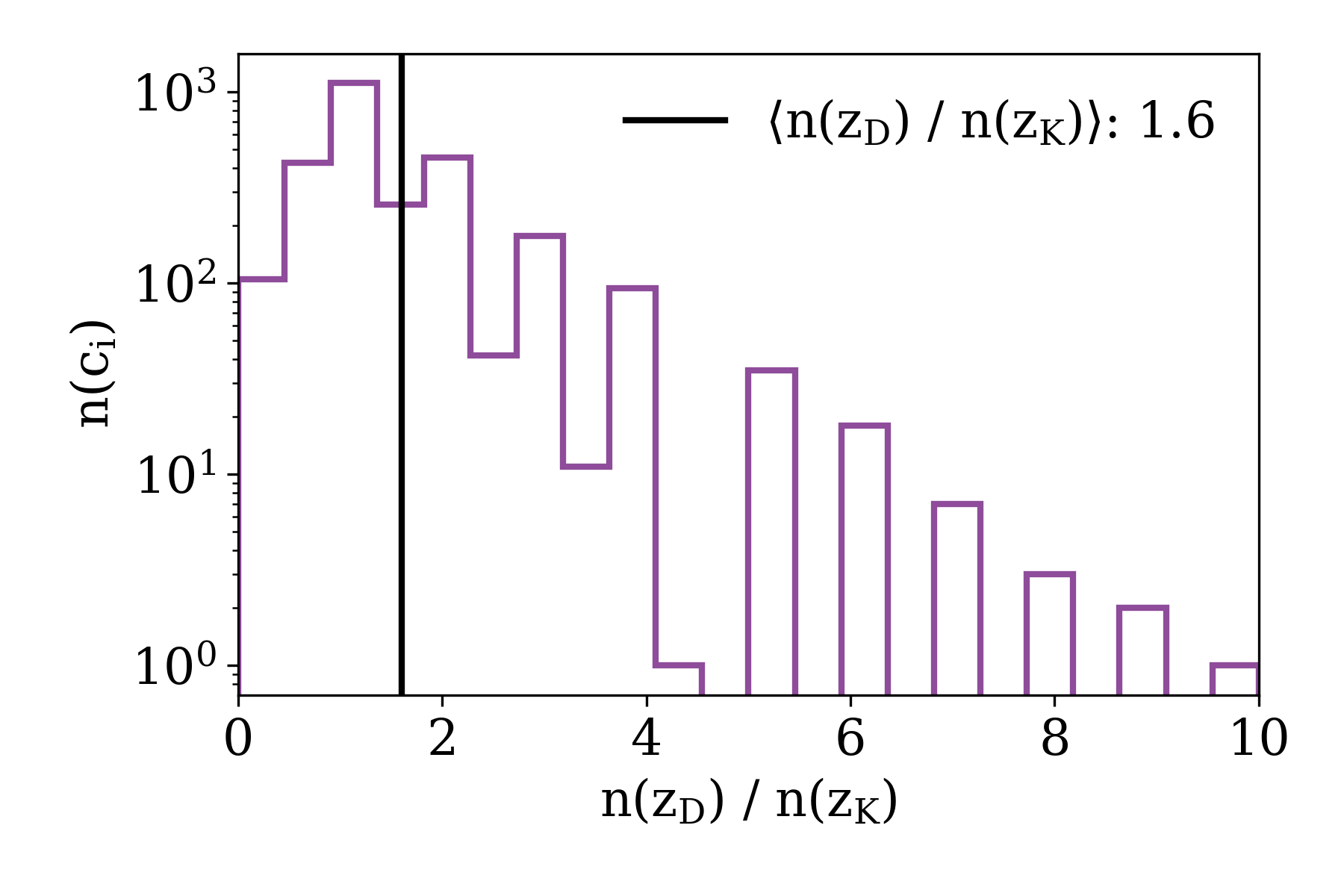}
\caption{Distribution of the relative unweighed occupation of shared SOM cells, n(c$_i$), where each cell is jointly occupied by both the DESI and KiDS COSMOS spectroscopic redshifts. This ratio is defined as the number of DESI spec-zs divided by the number of KiDS spec-zs per shared cell. On average, DESI provides 1.6 more redshifts than KiDS, and this is denoted by the black vertical line.}
\label{fig:relative_occupation_per_cell}
\end{figure}

Figure~\ref{fig:relative_occupation_per_cell} depicts the distribution of ratios of the DESI and KiDS spec-zs per Joint SOM cell. Across these Joint S$_K$ and Joint S$_D$ cells, the DESI spec-zs outnumber the KiDS calibration spec-zs by an average factor of 1.6, with some areas showing more than a five-fold increase. Notably, unique SOM cells populated exclusively by the DESI spec-zs are not included in this analysis. These DESI-only cells total 1,683, accounting for 16.5\% of the total SOM map. In addition, cells containing KiDS spec-z but no DESI spec-z account for 1,452 cells, corresponding to 14.2\% of the entire SOM. 
When restricting to the Joint SOM cells, 75\% of the KiDS photometric galaxies in SOM regions that can be calibrated are retained, while 25\% are removed.

\begin{figure*}
\centering
\includegraphics[width=0.7\textwidth, trim=0cm 1cm 0cm 1cm]{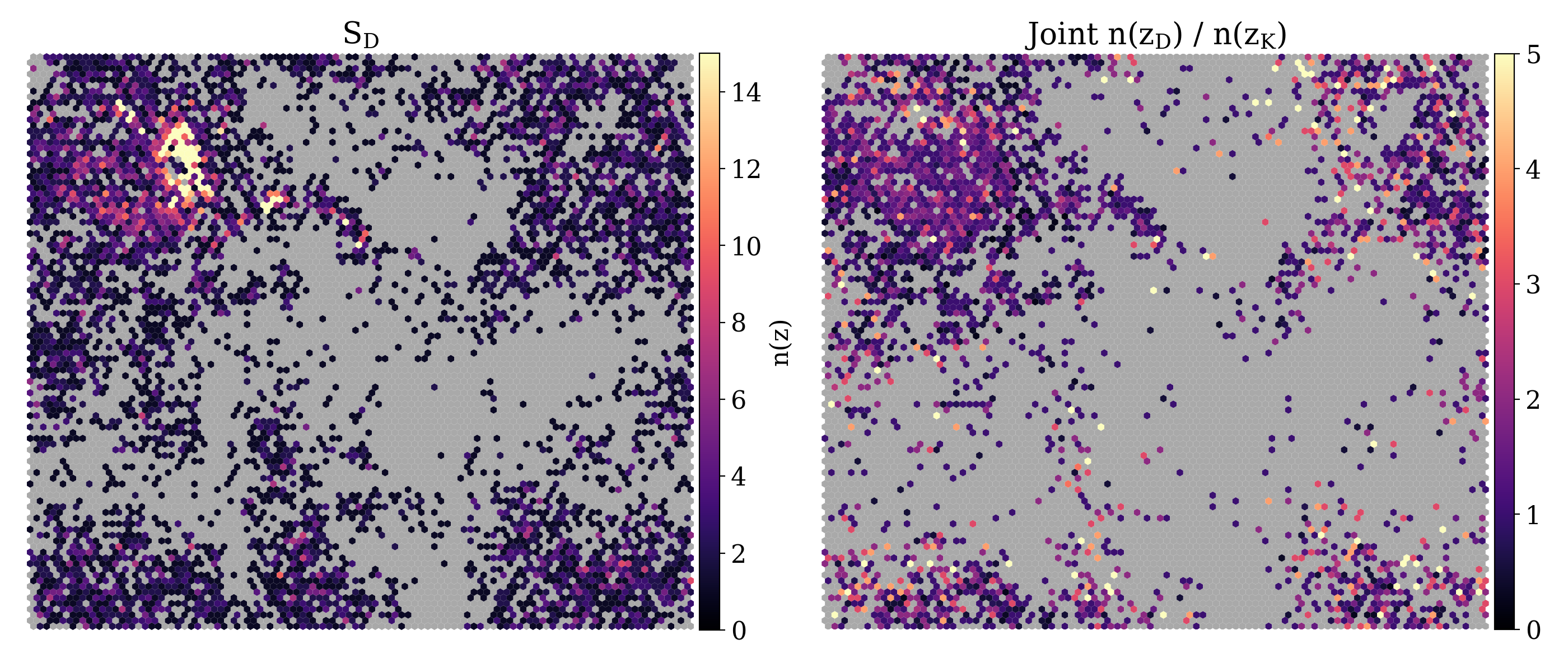}
\caption{   
Spec-z occupation and relative joint completeness in SOM space. 
The left panel shows the occupation density of the original $S_D$ sample 
(i.e., all DESI targets that pass the ~\citealt{Ratajczak2025} cuts), with colors representing the 
number of spectra in each SOM cell. Grey cells in the left panel correspond to regions unoccupied by $S_D$ (i.e., $n(z_D)=0$). The right panel shows the ratio of 
occupation between the joint $S_D$ and joint $S_K$ samples, where brighter 
colors indicate cells with a substantially higher count in DESI relative to KiDS. 
Grey cells in the right panel mark regions that are not jointly occupied by both surveys. The original $S_D$ sample covers 43.2\% of 
the entire SOM map. The left figure only includes the joint SOM cells, which are occupied by both KiDS and DESI, collectively accounting for 26.9\% of the entire SOM map.}
\label{fig:desi_occ_and_joint_kids_vs_desi_occ_ratio}
\end{figure*}

In Figure~\ref{fig:desi_occ_and_joint_kids_vs_desi_occ_ratio} we show the spectroscopic occupation of DESI in SOM space alongside the relative completeness of DESI and KiDS in their shared color–magnitude regions. The left panel displays the occupation density of the full $S_D$ sample, with color indicating the number of DESI spectra in each SOM cell. The right panel shows the ratio of occupation between the joint $S_D$ and joint $S_K$ samples, where brighter colors correspond to cells with a substantially higher count of DESI spectra relative to KiDS. Cells in grey are unoccupied. The full $S_D$ sample covers 43.2\% of the SOM map, while the joint SOM cells, occupied by both DESI and KiDS, account for 26.9\% of the map. By restricting our analysis to the shared SOM cells, we remove differences that arise purely from the two surveys occupying different regions of color-magnitude space. Within the shared SOM space, DESI increases the number of spectroscopic galaxies per cell, improving the sampling of $p(z|c)$. Consequently, any differences between Joint S$_D$ and Joint S$_K$ reflect survey-dependent differences in the shape of $p(z|c)$, rather than differences in which SOM cells are included.

\begin{figure*}
\centering
\includegraphics[width=\textwidth, trim=0cm 5cm 0cm 5cm]{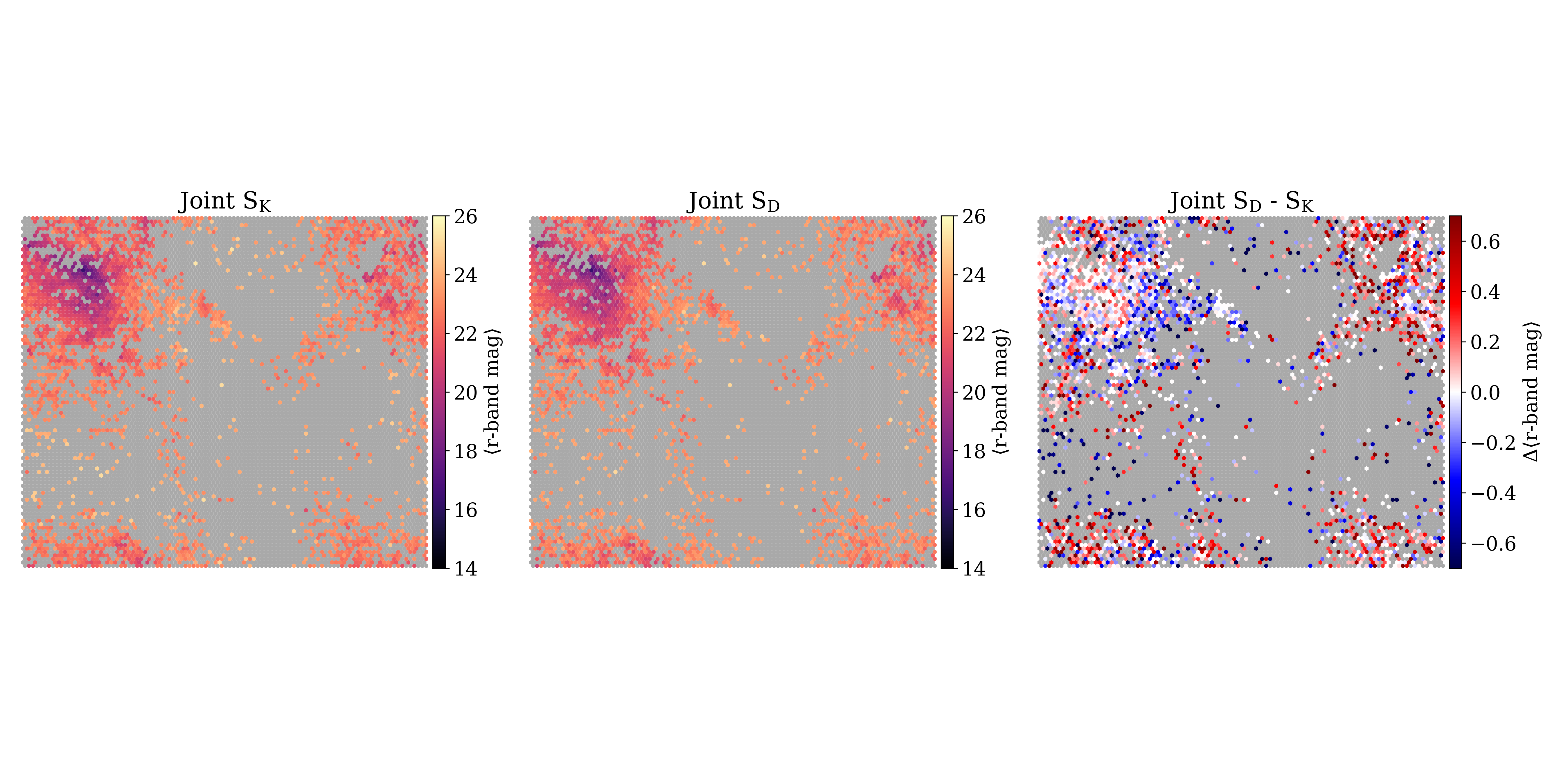}
\caption{Spatial distribution of the mean r-band magnitude within each joint SOM cell for the KiDS and DESI spectroscopic samples. The left panel shows the mean r-band magnitude for KiDS, calculated using KiDS deep photometry. The central panel illustrates the corresponding mean r-band magnitude for DESI, also derived from KiDS deep photometry. The right panel represents the magnitude difference between KiDS and DESI, where positive values indicate regions where KiDS appears relatively brighter (or DESI appears fainter). Some regions in SOM space exhibit notably large magnitude differences in absolute values, with values reaching up to~$\sim$0.6 magnitudes (highlighted in red or blue). Across all joint SOM cells, the average difference in $r$-band magnitude between DESI and KiDS is 0.24.}
\label{fig:kids_vs_desi_mean_mag_deep}
\end{figure*}

\begin{figure*}
\centering
\includegraphics[width=\textwidth, trim=0cm 5cm 0cm 5cm]{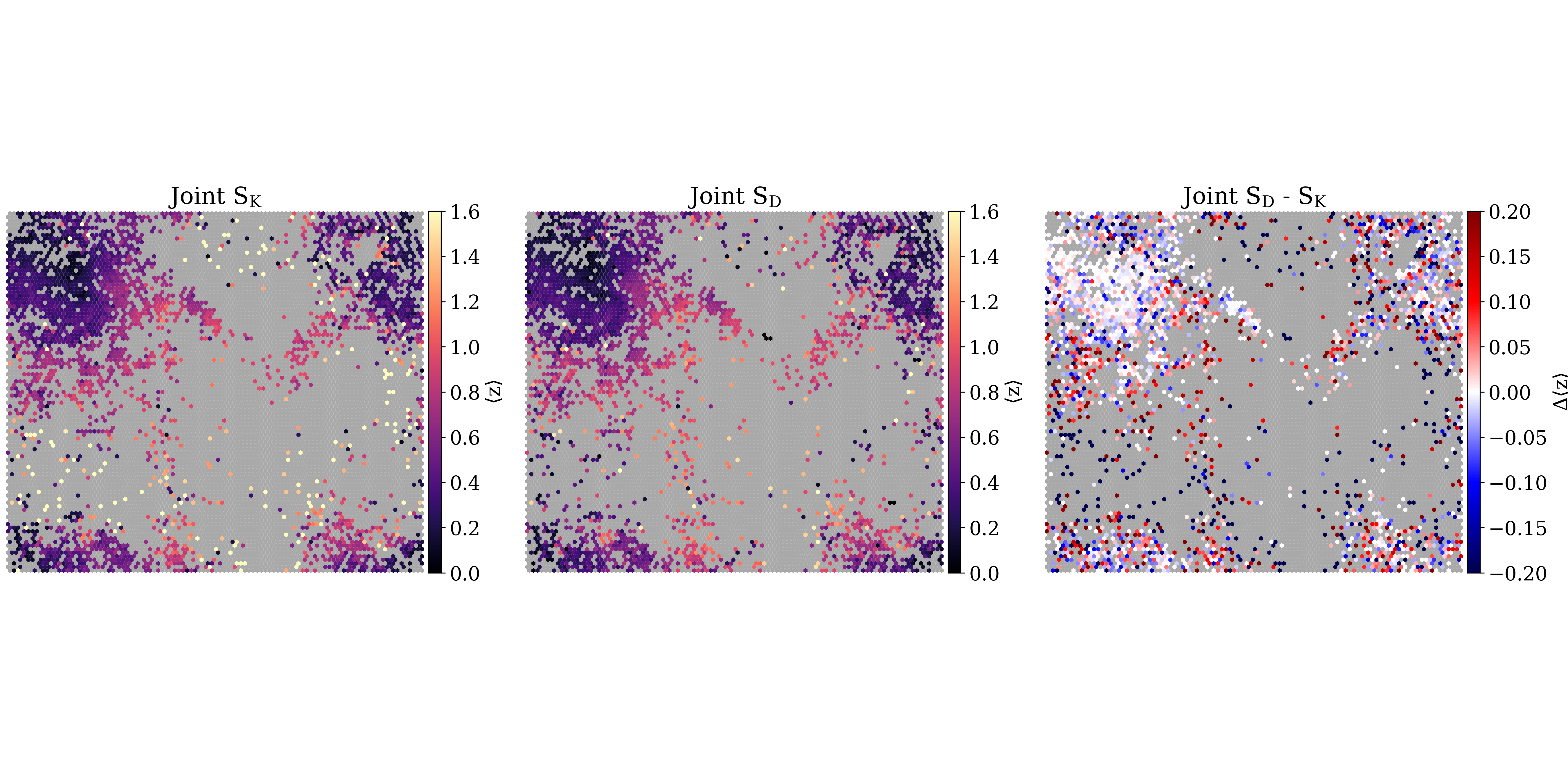}
\caption{Mean redshift per joint SOM cell. The leftmost panel illustrates the mean redshift per joint cell derived from KiDS data. The center panel presents the corresponding mean redshift for DESI. The rightmost panel quantifies the redshift difference between KiDS and DESI, where color intensity reflects the magnitude of the differences, with darker colors denoting cells where DESI reports lower redshifts. Regions with large discrepancies in mean redshifts correspond to areas of high incompleteness for DESI (see Figure~\ref{fig:desi_success}). This discrepancy arises because DESI is predominantly limited to $z \lesssim 1.6$, whereas KiDS incorporates the zCOSMOS-D compilation, which includes fainter, higher-redshift galaxies.}
\label{fig:kids_vs_desi_mean_and_differences}
\end{figure*}

\begin{figure*}
\centering
\includegraphics[width=0.8\textwidth]{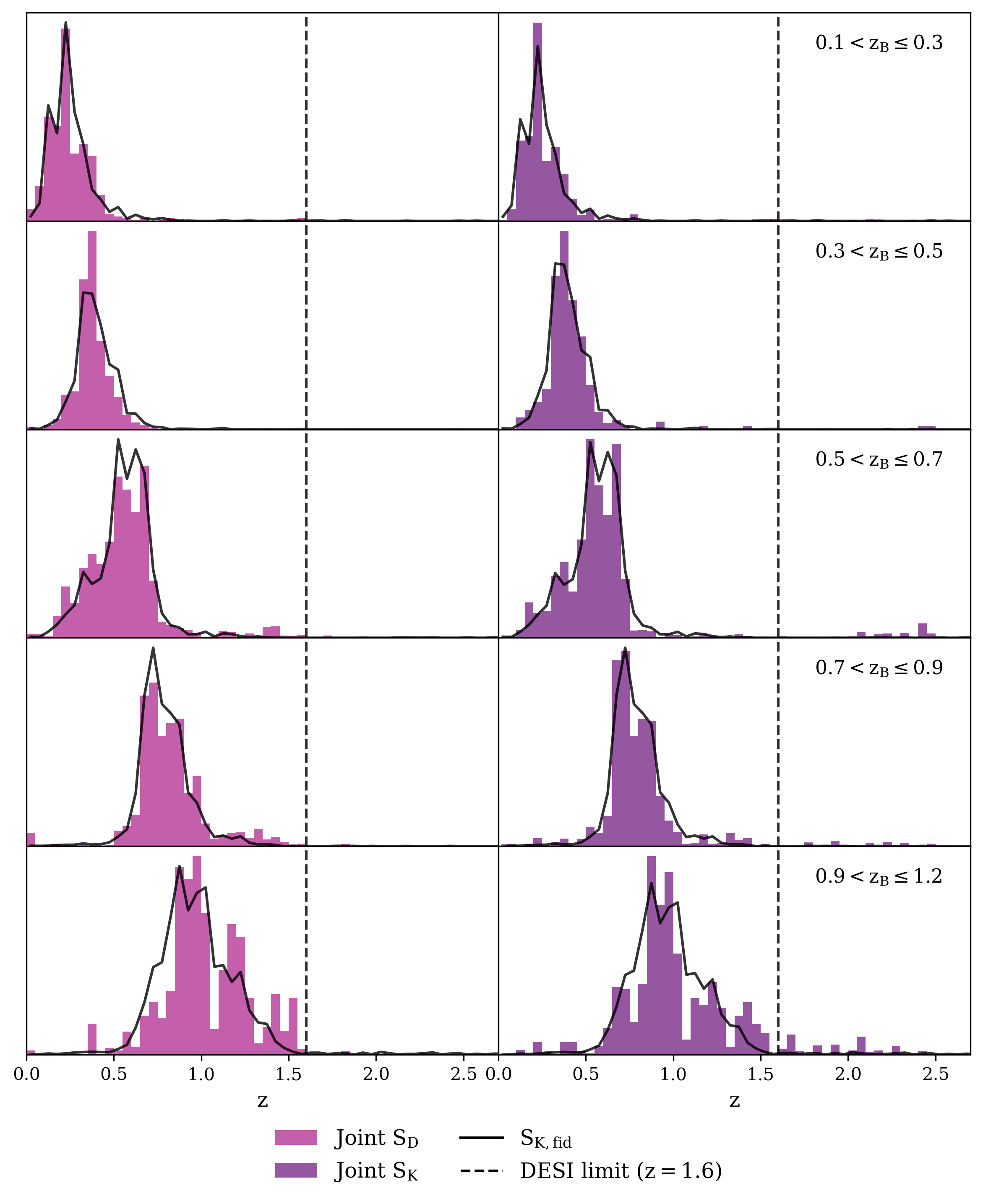}
\caption{The redshift distributions of Joint S$_D$ and Joint S$_K$ for the KiDS tomographic bins are shown, with Joint S$_D$ (pink) in the first column and Joint S$_K$ (purple) in the second column. The fiducial KiDS redshift distribution (S$_K$, fid) is in black in each bin for reference, and the DESI spectroscopic limit ($z = 1.6$) is denoted by a dashed vertical line. The third column summarizes the $\langle z \rangle$ and median $\tilde{z}$ values for each redshift distribution. The first ($0.1 < z \leq 0.3$) and last ($0.9 < z \leq 1.2$) tomographic bins show the largest shifts after restricting to joint SOM cells, with the first bin shifting towards lower redshifts and the last bin shifting towards higher redshifts. This suggests that the restriction to joint SOM cells most strongly impacts the lowest- and highest-z sources. In contrast, the intermediate bins ($0.3 < z \leq 0.9$) show smaller deviations, with Joint S$_D$ closely following S$_K$, fid. Across most bins, Joint S$_K$ consistently has a slightly higher $\langle z \rangle$ than Joint S$_D$. The exact values for $\langle z\rangle$, $\tilde{z}$, $\Delta\langle z\rangle$, and $\Delta\tilde{z}$ are reported in Table~\ref{tab:joint_means_medians}.
}
\label{fig:mean_median_nz}
\end{figure*}

In Figure~\ref{fig:kids_vs_desi_mean_mag_deep}, we present the r-band magnitude distribution within the Joint SOM cells shared by KiDS and DESI, both derived from KiDS deep photometry. The left panel visualizes the r-band magnitude distribution for Joint S$_K$, while the middle panel displays the equivalent for Joint S$_D$. Most notably, we quantify the r-band magnitude difference between the two datasets in the right panel, restricted to their shared SOM cells. This comparison highlights variations in observed depth, despite both datasets using KiDS photometry. Because the SOM is defined in color–magnitude space, each cell spans a limited magnitude range; the differences shown therefore reflect selection differences between the DESI and KiDS spectroscopic samples within matched color–magnitude regions. These differences arise from the distinct galaxy selection criteria in the spec-z samples of DESI and KiDS. Although there are regions in Joint SOM space where DESI shows significantly higher magnitudes (up to $\sim$0.6, highlighted in yellow), the overall r-band magnitude difference, calculated as the simple arithmetic mean across all Joint SOM cells, is 0.24, indicating that DESI provides, on average, a slightly fainter sample than the KiDS calibration set.

Figure~\ref{fig:kids_vs_desi_mean_and_differences} presents a comparison of the mean redshifts within the Joint SOM cells. In the Joint S\(_K\) calibration sample, the highest-redshift SOM cells (indicated by the bright yellow colors, corresponding to \(z \sim 1.6\)) are clearly visible across several regions of the map. In contrast, the central panel representing the Joint S\(_D\) sample shows a similar overall structure but systematically lower redshifts in these same cells. This discrepancy reflects the DESI redshift limit of \(z < 1.6\), where the [O\,\textsc{ii}] emission line exits the DESI wavelength range, reducing DESI’s ability to recover high-redshift galaxies in these cells. The variation in mean redshifts between Joint S\(_K\) and Joint S\(_D\) is examined in more detail in Section~\ref{subsec:comp}.

\subsection{Comparison of the KiDS and DESI $n(z)$, $\langle$z$\rangle$, and $\tilde{z}$}

We now derive $n(z)$ distributions using the Joint S$_D$ calibration sample instead of the original KiDS calibration catalog (it is important to note that these do correspond to the $n(z)$ of the published KiDS source distributions). Figure~\ref{fig:mean_median_nz} presents the comparison of the redshift distributions for Joint S$_D$ (pink) and Joint S$_K$ (purple), alongside the fiducial KiDS $n(z)$ (black) for reference.

In the joint SOM cells, KiDS and DESI provide two independent empirical estimates of the conditional redshift distribution $p(z|c)$. Differences between the resulting $n(z)$s can therefore be interpreted as shifts in the survey-specific relations $p(z_K|c)$ and $p(z_D|c)$, reflecting how each survey samples the underlying galaxy population in color-magnitude space. Because SOMPZ relies on an accurate reconstruction of $p(z|c)$ to build $n(z)$, quantifying where KiDS and DESI differ is essential for assessing the resulting calibration.

Our results show agreement between Joint S$_D$ and Joint S$_K$ in the intermediate tomographic bin, while larger shifts appear at the lowest and highest redshifts. As shown in Figure~\ref{fig:mean_median_nz} and Table~\ref{tab:joint_means_medians}, restricting KiDS to the Joint SOM cells shifts the first tomographic bin toward lower redshifts and the highest bin toward higher redshifts, indicating that the restriction primarily impacts the edges of the redshift distribution. This pattern can be interpreted through the lens of $p(z|c)$ and $p(c|z)$: because the DESI spectroscopic sample is systematically fainter, DESI and KiDS populate SOM cells with different $p(c|z)$ even at the same true redshift. Limiting the calibration to the Joint SOM cells therefore samples low- and high-$z$ populations differently in the two surveys, shifting the inferred $p(z|c)$, and thus the mean and median $n(z)$, in the lowest and highest tomographic bins, while leaving the intermediate bin largely unchanged.

Outliers in this context refer to SOM cells whose calibrated mean redshifts lie outside of the expected distribution.
These manifest as tails in the $n(z)$s, particularly at higher redshifts, which pull the mean and median redshifts toward higher values. The mean redshift is more sensitive to extreme values, while the median is more robust to catastrophic failures, regions in SOM space where photo-z and spec-z measurements strongly disagree. The KiDS team has previously noted that their conservative spectroscopic flagging in higher tomographic bins reduces the spectroscopic representation of high-z sources, meaning fewer high-z galaxies are retained in the calibration sample relative to the underlying population. In contrast, when we restrict the analysis to Joint SOM cells, we preferentially select regions of color-magnitude space that are well sampled by both KiDS and DESI spectroscopy, which tends to increase the fraction of high-z galaxies represented in the calibration sample. This leads to slightly higher mean and median redshifts relative to the fiducial KiDS $n(z)$.

The difference in spectroscopic coverage likely plays a role in the discrepancies observed in our calibration, particularly as DESI is not optimized to detect high-redshift galaxies at the same depths as zCOSMOS-D, which includes a more diverse sample of faint, high-z sources.

\begin{table}
    \centering
    \begin{tabular}{c l rr}
    \hline\hline
    Tomographic Bin & Sample & $\langle z\rangle$ & $\tilde{z}$ \\
    \hline
    \multirow{4}{*}{$0.1 < z_B \leq 0.3$}
      & $S_{K,\mathrm{fid}}$      & 0.257 & 0.234 \\
      & Joint $S_D$               & 0.258 & 0.221 \\
      & Joint $S_K$               & 0.293 & 0.222 \\
      & Joint$_{S_D-S_K}$         & $-0.035$ & $-0.001$ \\
    \hline
    \multirow{4}{*}{$0.3 < z_B \leq 0.5$}
      & $S_{K,\mathrm{fid}}$      & 0.403 & 0.385 \\
      & Joint $S_D$               & 0.376 & 0.365 \\
      & Joint $S_K$               & 0.426 & 0.374 \\
      & Joint$_{S_D-S_K}$         & $-0.049$ & $-0.009$ \\
    \hline
    \multirow{4}{*}{$0.5 < z_B \leq 0.7$}
      & $S_{K,\mathrm{fid}}$      & 0.564 & 0.568 \\
      & Joint $S_D$               & 0.560 & 0.551 \\
      & Joint $S_K$               & 0.603 & 0.553 \\
      & Joint$_{S_D-S_K}$         & $-0.043$ & $-0.002$ \\
    \hline
    \multirow{4}{*}{$0.7 < z_B \leq 0.9$}
      & $S_{K,\mathrm{fid}}$      & 0.792 & 0.772 \\
      & Joint $S_D$               & 0.821 & 0.794 \\
      & Joint $S_K$               & 0.817 & 0.754 \\
      & Joint$_{S_D-S_K}$         & $+0.003$ & $+0.040$ \\
    \hline
    \multirow{4}{*}{$0.9 < z_B \leq 1.2$}
      & $S_{K,\mathrm{fid}}$      & 0.984 & 0.958 \\
      & Joint $S_D$               & 1.026 & 0.985 \\
      & Joint $S_K$               & 1.040 & 0.957 \\
      & Joint$_{S_D-S_K}$         & $-0.013$ & $+0.028$ \\
    \hline
    \end{tabular}
    \caption{
    Mean redshift $\langle z\rangle$ and median redshift $\tilde{z}$ for the fiducial KiDS calibration sample $S_{K,\mathrm{fid}}$, the Joint $S_D$ and Joint $S_K$ calibration samples, and their differences Joint$_{S_D-S_K}$ in each tomographic bin.}
    \label{tab:joint_means_medians}
\end{table}

\begin{figure}
\centering
\includegraphics[width=\columnwidth, trim=0cm 0cm 0cm 0cm]{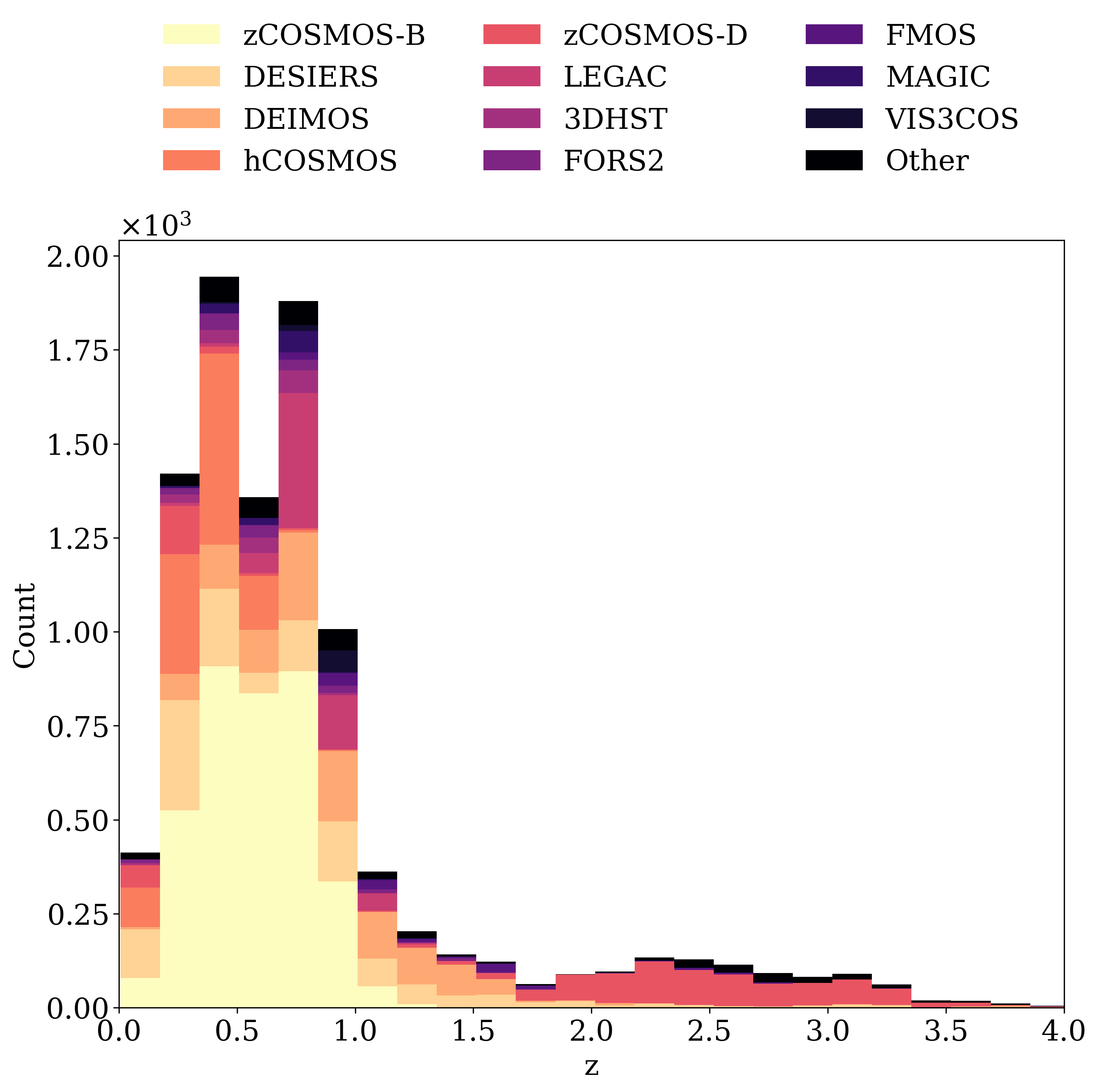}
\caption{Distribution of the different spec-z samples in the ``COSMOS Team'' compilation of KiDS COSMOS redshifts. Each color represents a different survey contributing to the compilation, with zCOSMOS-D shown in coral. The histogram illustrates the redshift distribution (z) of the combined dataset, highlighting the relative contributions of each survey across the redshift range.}
\label{fig:zcosmos_all}
\end{figure}

\begin{figure}
\centering
\includegraphics[width=\columnwidth, trim=0cm 0cm 0cm 0cm]{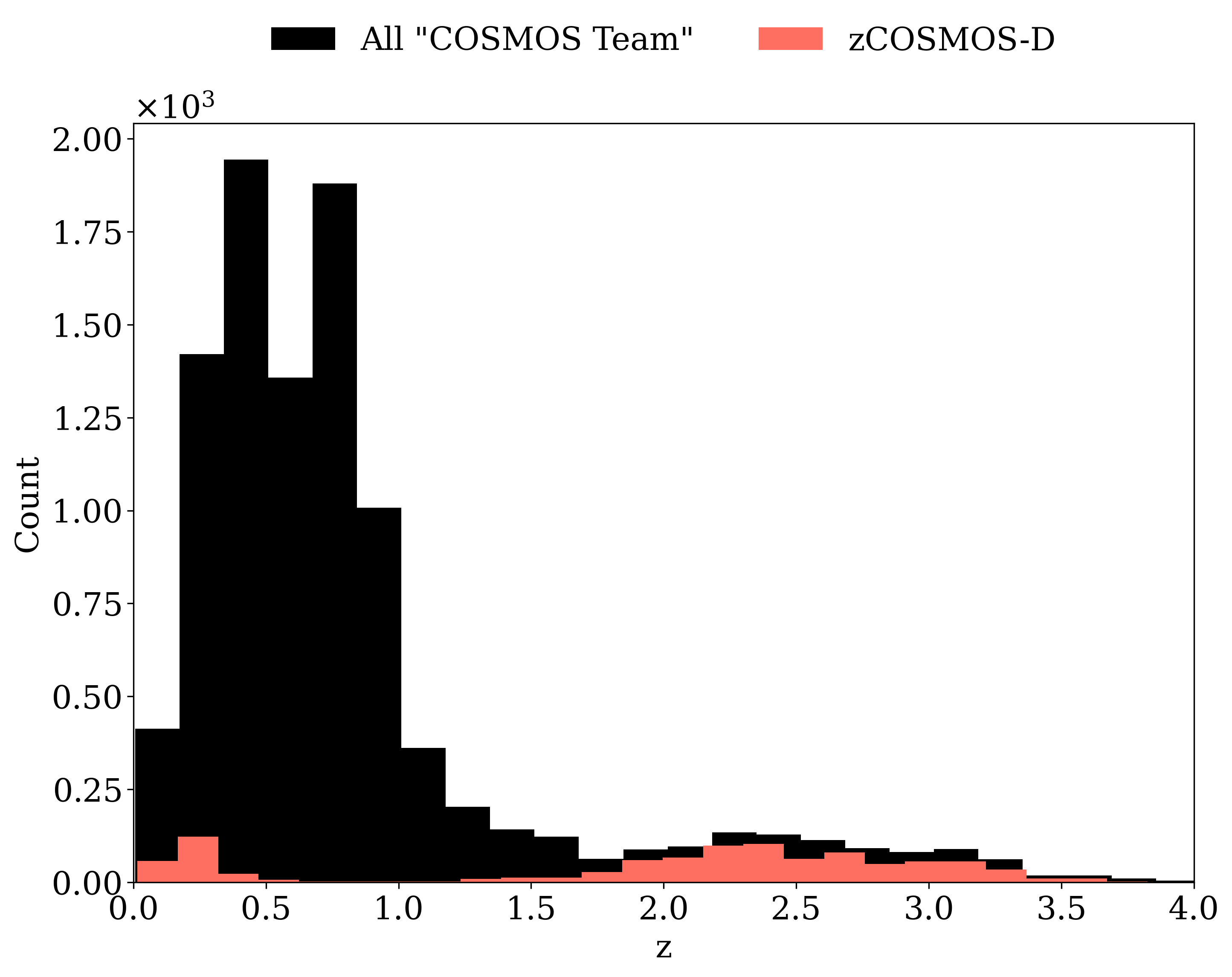}
\caption{Redshift distribution of the COSMOS Team compilation (in black) and zCOSMOS-D (in coral), similar to the previous figure. The black histogram represents the combined COSMOS Team redshifts, excluding zCOSMOS-D, while the coral highlights the zCOSMOS-D contribution. Notably, zCOSMOS-D occupies the higher redshift range, particularly above the DESI redshift limit of z $>$ 1.6, extending beyond the DESI spectral detection range. zCOSMOS-D sample in covers the redshift range inaccessible to DESI.}
\label{fig:zcosmos_d}
\end{figure}

\begin{figure*}
\centering
\includegraphics[width=0.7\textwidth, trim=0cm 0cm 0cm 0cm]{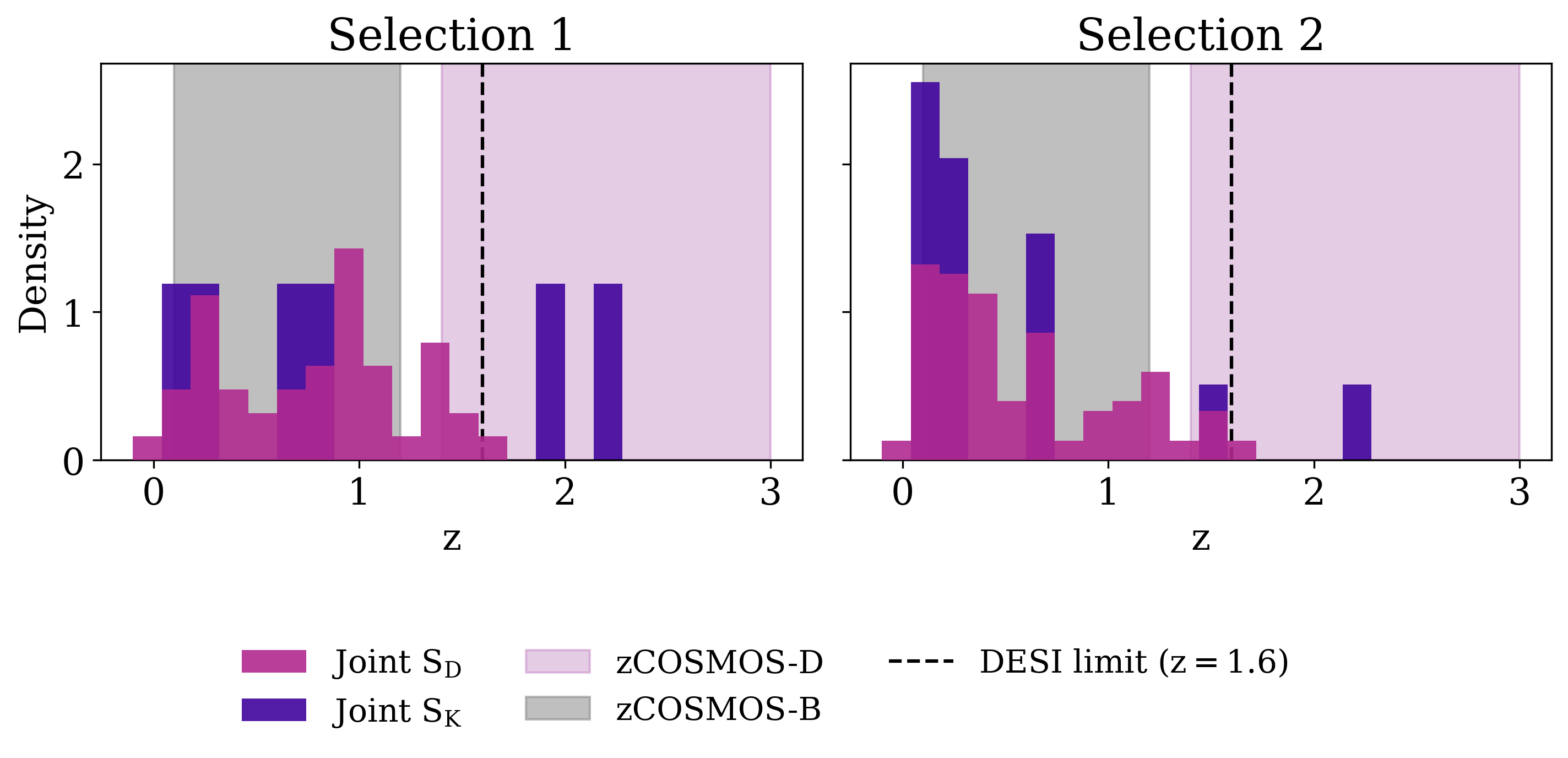}
\caption{Redshift distribution of the two selections as shown in Figure~\ref{fig:desi_success}. The pink represents the Joint S$_D$ z distribution, while the purple depicts the Joint S$_K$ distribution. The dotted black line indicates the DESI z limit at $z = 1.6$. The grey and purple shaded regions correspond to the z ranges of zCOSMOS-B and zCOSMOS-D, respectively. It is evident that the KiDS COSMOS compilation of spec-z’s extends beyond the z range covered by DESI. These selections align with regions of significant discrepancies in color-space between KiDS and DESI.}
\label{fig:desi_vs_kids_zcosmos}
\end{figure*}

\begin{figure}
\centering
\includegraphics[width=\columnwidth, trim=0cm 5.5cm 0cm 5.5cm]{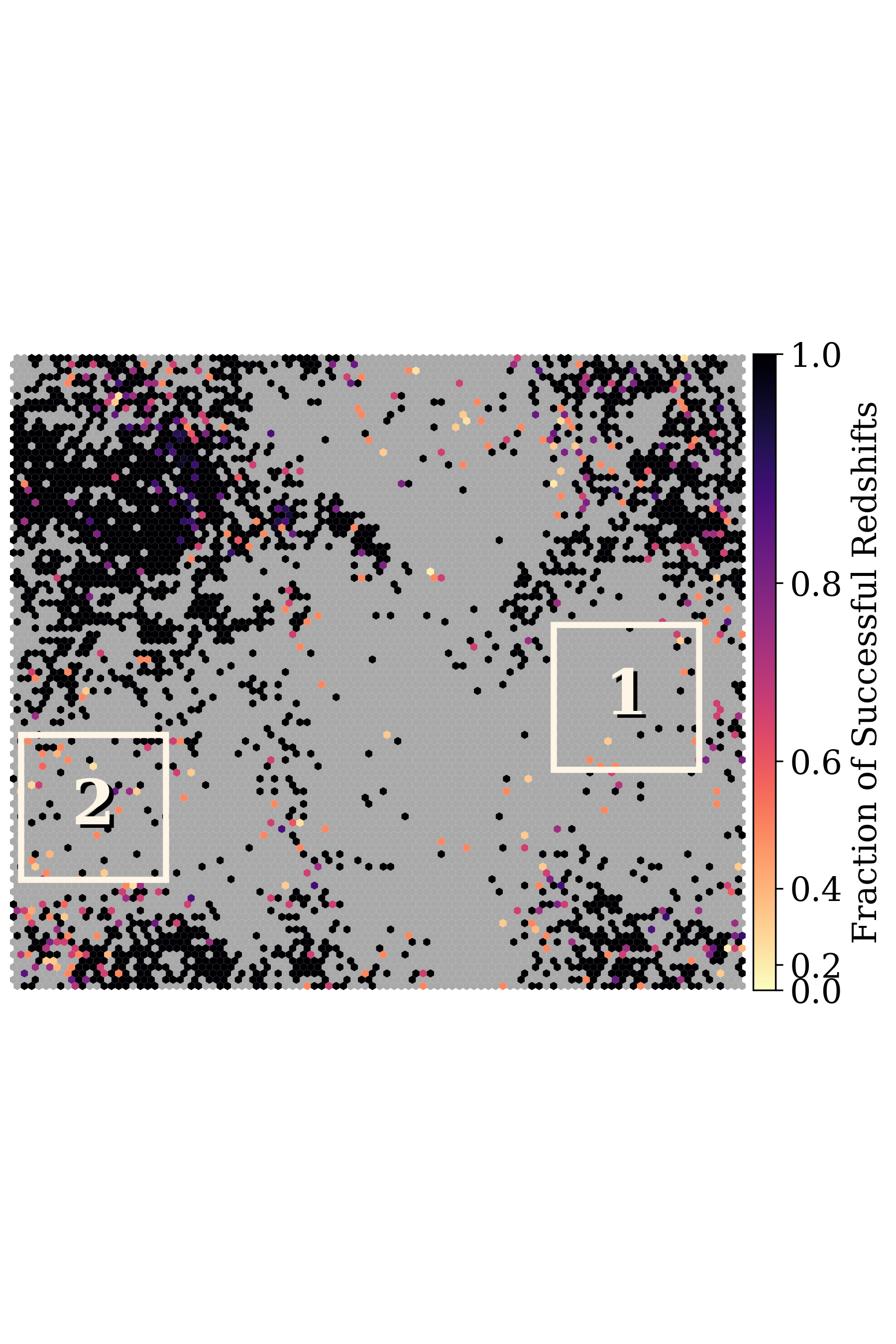}
\caption{The fraction of successful redshift measurements in Joint S$_D$, defined following \citet{Ratajczak2025}. Darker colors indicate SOM cells with higher DESI success rates. The two highlighted square regions were selected by identifying contiguous $20 \times 20$ patches that exhibit the most negative average $\Delta\langle z\rangle$ between Joint S$_K$ and Joint S$_D$ (see Fig.~\ref{fig:kids_vs_desi_mean_and_differences}). These regions correspond to areas where DESI systematically reports lower mean redshifts than KiDS and where the DESI success rate is comparatively low. The squared regions provide a compact and visually coherent way to compare success-rate structure in areas with the strongest discrepancies between the two surveys.}
\label{fig:desi_success}
\end{figure}

\begin{figure}
\centering
\includegraphics[width=0.9\columnwidth, trim=0cm 0cm 0cm 0cm]{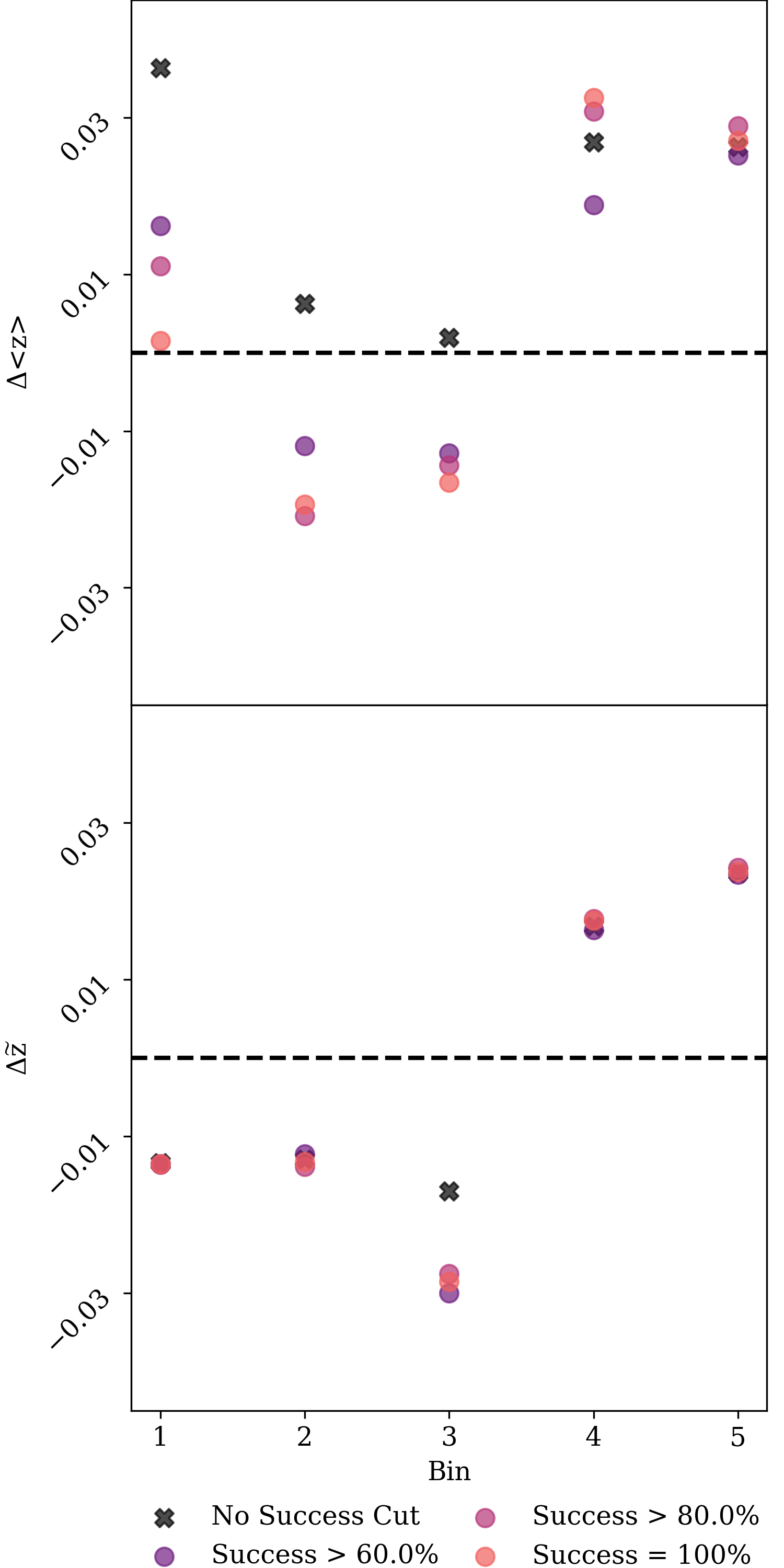}
\caption{
Mean (top row) and median (bottom row) redshift differences between Joint S$_D$ and the fiducial KiDS redshift distributions S$_{K,\mathrm{fid}}$, shown as a function of the completeness threshold applied to the DESI calibration sample. Completeness is defined following \citet{Ratajczak2025}. The color of each dot corresponds to a specific DESI completeness level per color cell: black crosses indicate no completeness cuts, while purple, magenta, and orange dots represent completeness levels of 60\%, 80\%, and 100\%, respectively.
}
\label{fig:completness_diff_mean_z}
\end{figure}

\begin{figure*}
\centering
\includegraphics[width=\textwidth, trim=0cm 0cm 0cm 0cm]{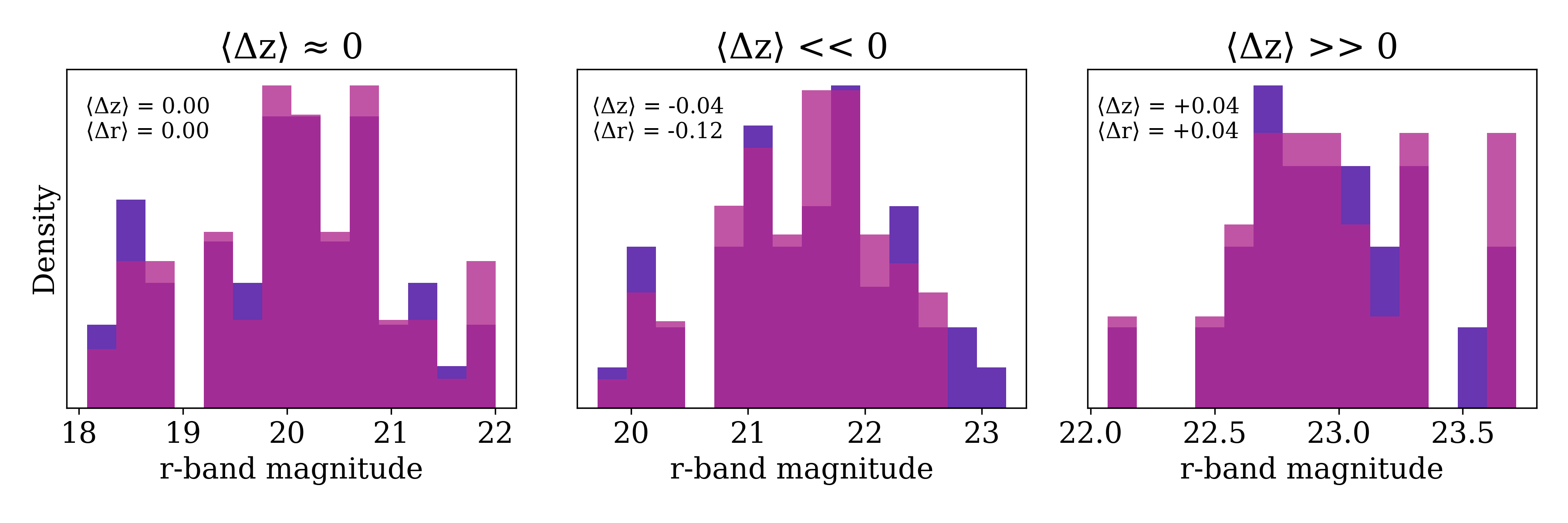}
\caption{Stacked $r$-band magnitude distributions for Joint S$_D$ (pink) and Joint S$_K$ (purple) in three representative sets of joint SOM cells. Each panel combines several cells selected to have (left) high DESI success rates and $\Delta\langle z\rangle \approx 0$, (middle) low DESI success rates and $\Delta\langle z\rangle \ll 0$, and (right) low DESI success rates and $\Delta\langle z\rangle \gg 0$, where $\Delta\langle z\rangle = \langle z\rangle_{\rm Joint\,S_D} - \langle z\rangle_{\rm Joint\,S_K}$. In the $\Delta\langle z\rangle \approx 0$ case, the DESI and KiDS magnitude distributions closely agree and $\Delta\langle r\rangle \approx 0$. In the $\Delta\langle z\rangle \ll 0$ group, the DESI sample is systematically brighter ($\Delta\langle r\rangle < 0$), consistent with DESI missing faint, high-redshift galaxies relative to KiDS in those cells. In the $\Delta\langle z\rangle \gg 0$ group, the DESI sample is slightly fainter on average ($\Delta\langle r\rangle > 0$).
}
\label{fig:cell_mag_hists}
\label{fig:desi_target_types}
\end{figure*}

\begin{figure}
\centering
\includegraphics[width=\columnwidth, trim=5cm 5cm 5cm 0cm]{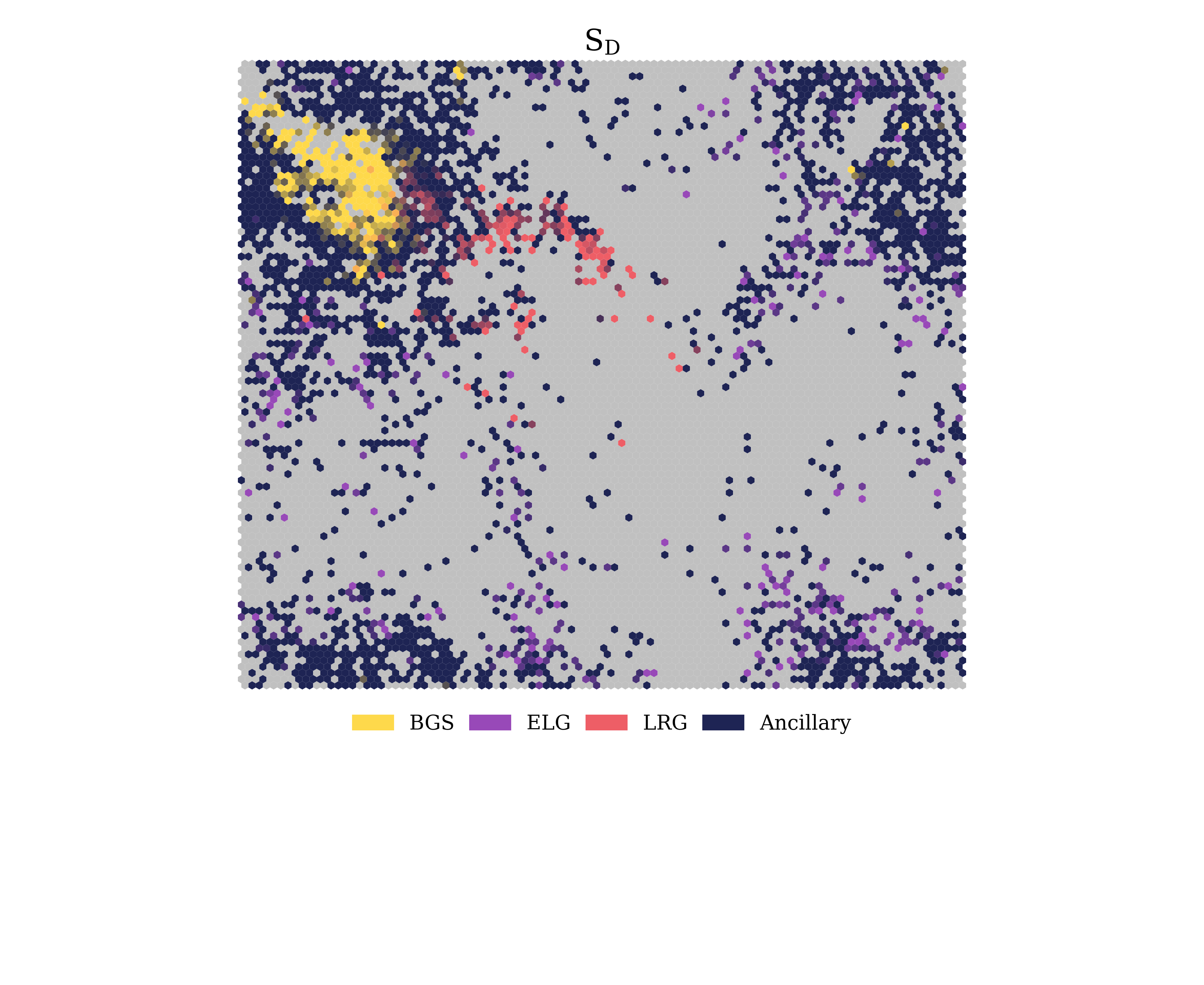}
\caption{The relative distribution of different DESI targets within the shared SOM space of Joint DESI cells. The colors represent BGS (Bright Galaxies, yellow), ELGs (Emission Line Galaxies, purple), LRGs (Luminous Red Galaxies, coral), and Ancillary targets (dark blue), and have been scaled to enhance visibility. The lower portion of the SOM map shows a higher fraction of ELG-dominated cells. These ELG-dominated cells correspond to regions where DESI experiences difficulty acquiring accurate redshift measurements.}
\label{fig:desi_target_types}
\end{figure}

\diana{Using $\Delta\langle z\rangle/(1+\langle z\rangle_{\rm Joint,S_K})$, only bins 1 and 3 meet the LSST DESC Science Requirements Document (SRD) redshift-bias target for Year 1 (Y1),
$\frac{|\Delta\langle z\rangle|}{1+\langle z\rangle} < 0.002$,
and none meet the more stringent Year 10 (Y10) requirement,
$\frac{|\Delta\langle z\rangle|}{1+\langle z\rangle} < 0.001$.
The largest offsets at high $z$ reach $\sim 0.009$, implying additional calibration is required to keep redshift systematics subdominant \citep{DESCScienceRoadmap2021}.}

While this work focuses on a calibration sample restricted to Joint SOM cells and spectroscopic redshifts, a complementary approach is explored in our companion paper~\citep{lange2025}, which incorporates a hybrid of photo-z and spec-z measurements to expand coverage across the full color–magnitude space. This alternative method is particularly useful for supplementing regions of color space that are underrepresented in spectroscopic datasets like DESI's. 
Although the inclusion of photo-zs means that the resulting calibration is not strictly comparable to our purely spectroscopic-based approach since the inferred $p(z|c)$ incorporates photo-z assumptions, it nevertheless offers a promising avenue for future weak lensing surveys. Further work will be needed to assess the influence of using the COSMOS 30-band photo-z catalog in these hybrid calibration strategies.

\subsection{Spectroscopic Sample Differences}

In Figures~\ref{fig:zcosmos_all} and~\ref{fig:zcosmos_d}, we show the distribution of spec-z samples in the KiDS ``COSMOS Team'' compilation. The histograms highlight the contributions of each survey, with zCOSMOS-D appearing in coral. In Figure~\ref{fig:zcosmos_d}, the black histogram shows the COSMOS Team spec-zs (excluding zCOSMOS-D), while the coral highlights zCOSMOS-D, which extends into the higher redshift range, particularly above the DESI limit of z~$<$~1.6. This allows zCOSMOS-D and thus KiDS-1000 to cover regions inaccessible to DESI.

The inclusion of the zCOSMOS-D sample by KiDS, may also contribute to the observed discrepancies in the mean redshift of targets within the SOM space. Figure \ref{fig:kids_vs_desi_mean_and_differences} indicates a disparity in the redshifts across certain regions of color-space. As previously discussed, DESI is constrained by a redshift limit of z~$<$~1.6, which hampers its ability to detect higher redshifts. Conversely, the KiDS COSMOS Team compilation encompasses the zCOSMOS-D survey, which extends beyond the DESI limit. This difference in high-z spectroscopic reach may account for the observed variations.

The zCOSMOS sample was selected using a combination of $U$, $B$, $V$, and $R$ cuts along with a ``BzK'' selection \citep{Lilly2009_ApJS_184_218}. In addition, redshift success requires the presence of ultraviolet absorption features. In future work, it would be valuable to understand whether these specific selections lead to a biased population sample when compared to other galaxies that populate the same color magnitude cells. This would be especially relevant as zCOSMOS-Deep contributes galaxies both at low and high redshifts and are the primary source contributing to the high-z tails.

However, the unexpectedly higher mean redshift of DESI in certain regions of the SOM may result from its target selection and sample size. First, the DESI spectroscopic sample prioritizes specific galaxy populations, such as luminous red galaxies (LRGs) and emission-line galaxies (ELGs), which are more concentrated at intermediate redshifts (z $\sim$~0.7 to 1.2). This focus can bias the mean redshift higher in certain SOM regions compared to the broader spectroscopic samples included in KiDS, which capture a wider range of galaxy types and lower-redshift sources. Second, the DESI sample in the shared KiDS footprint and SOM cells is significantly larger (8,473 spec-zs compared to 6,266 for KiDS), increasing the likelihood that these regions will contain more high-redshift galaxies simply by chance, e.g., by sampling clustered structures or overdensities at those redshifts, an effect related to cosmic variance.

Having observed the persistent shift in the redshift distribution, we next turn our focus to examining the impact of spectroscopic incompleteness on these $n(z)$ distributions.

\subsection{Impact of Spectroscopic Coverage and Completeness on Redshift Distributions}
\label{subsec:comp}

Figure~\ref{fig:desi_vs_kids_zcosmos} provides a comparison of the spec-z distributions from Joint S$_D$ and Joint S$_K$ in two selected regions of color-space where the redshift distributions exhibit significant discrepancies. These regions are particularly interesting because they correspond to areas where the spectroscopic completeness of DESI is lower, suggesting that differences in spectroscopic selection could be influencing the observed redshift distributions.

The KiDS COSMOS compilation of spec-z’s extends beyond the redshift range covered by Joint S$_D$, largely due to the capabilities of the zCOSMOS survey. This survey, conducted using the VIMOS spectrograph on the 10-meter Very Large Telescope (VLT), was split into two parts: zCOSMOS-B, which targeted $\sim$20k I-band selected galaxies at z~$<$~1.2, and zCOSMOS-D, which focused on a smaller sample of galaxies in the higher-redshift range 1.5~$<$~z~$<$~3.0. While zCOSMOS-B covers the full COSMOS field, zCOSMOS-D was designed specifically to probe high-redshift galaxies using specific color cuts described in the previous section.

\begin{table}
\centering
\begin{tabular}{lcc}
\hline\hline
Completeness Threshold 
    & Spec-z Kept 
    & Photo-z Kept \\
\hline
$> 60\%$ 
    & 94.5\% 
    & 95.3\% \\
$> 80\%$ 
    & 88.2\% 
    & 89.4\% \\
$= 100\%$ 
    & 73.2\% 
    & 81.0\% \\
\hline
\end{tabular}
\caption{
Fraction of galaxies retained after applying DESI success-rate completeness cuts.
Spec-z percentages refer to the Joint SOM calibration sample, and photo-z percentages refer to the KiDS source galaxies 
in calibratable SOM cells.}
\label{tab:completeness_kept}
\end{table}

To better understand the impact of the spectroscopic selection of DESI, we examine its redshift success rates. Figure~\ref{fig:desi_success} presents an overview of the success rate of spec-z measurements within Joint S$_D$, as determined by our selection criteria. The two highlighted square regions were chosen based on where the average $\Delta\langle z\rangle$ is the most negative. A key feature of this visualization is that the two selected regions correspond to areas where DESI exhibits lower success rates, further supporting the notion that spectroscopic completeness limitations contribute to the redshift discrepancies observed between Joint S$_D$ and Joint S$_K$.

In this work, we define a galaxy as ``successful'' if it passes the quality cuts outlined in Table~\ref{tab:desi_quality_cuts}. Completeness, as defined by \citet{Ratajczak2025}, is determined by the fraction of galaxies within a SOM cell that meet this success criterion. For example, a completeness level of 60\% corresponds to selecting only SOM cells where at least 60\% of galaxies pass these quality cuts. Figure~\ref{fig:completness_diff_mean_z} illustrates the extent to which variations in completeness levels affect the mean and median redshift in comparison to the fiducial KiDS redshift distributions, showing how increasing completeness impacts the recovered $n(z)$.

As the completeness level increases (e.g., from no cuts to 60\%, 80\%, and 100\%), galaxies in SOM cells with insufficient spectroscopic sampling are progressively removed. The corresponding fractions of spectroscopic and photometric galaxies retained at each completeness threshold are summarized in Table~\ref{tab:completeness_kept}. This removal changes the inferred ⟨z⟩ depending on the redshift structure of the low-completeness cells. In the higher tomographic bins, the low-completeness cells tend to contain a larger fraction of low-z galaxies. Dropping these cells therefore shifts ⟨z⟩ upwards. Conversely, in the lower tomographic bins, the low-completeness cells contain a large fraction of high-z galaxies, so removing them shifts ⟨z⟩ downward. 

From a probabilistic perspective, lowering the completeness threshold admits SOM cells in which $p(z|c)$ is poorly constrained by spectroscopic observations, while raising the threshold removes cells whose estimates of $p(z|c)$ are dominated by selection-driven differences in $p(c|z)$. These changes propagate directly into the inferred $n(z)$.

To further illustrate how spectroscopic selection effects couple depth and redshift in individual SOM cells, we select three representative sets of joint cells and compare their $r$-band magnitude distributions for Joint S$_D$ and Joint S$_K$. We first restrict to cells with reasonably well-sampled spectroscopy (requiring at least $\sim$7--10 DESI and KiDS spec-$z$ per cell and similar total counts from each survey) and then group cells according to their mean redshift offset $\Delta\langle z\rangle = \langle z\rangle_{\rm Joint\,S_D} - \langle z\rangle_{\rm Joint\,S_K}$ and DESI success fraction. The first group contains high-success cells with $\Delta\langle z\rangle \approx 0$, the second consists of low-success cells with $\Delta\langle z\rangle \ll 0$, and the third contains low-success cells with $\Delta\langle z\rangle \gg 0$. For each group, we stack the DESI and KiDS $r$-band magnitudes and show their normalized distributions (Fig.~\ref{fig:cell_mag_hists}). In the $\Delta\langle z\rangle \approx 0$ case, the DESI and KiDS magnitude distributions are nearly identical, and the corresponding mean $r$-band magnitudes also agree ($\Delta\langle r\rangle \approx 0$). By contrast, in cells where DESI systematically reports lower mean redshifts ($\Delta\langle z\rangle \ll 0$), the DESI calibration sample is systematically brighter than the KiDS sample ($\Delta\langle r\rangle < 0$), indicating that DESI preferentially misses faint, high-$z$ galaxies in these cells. Conversely, in cells with $\Delta\langle z\rangle \gg 0$, the DESI sample is slightly fainter on average ($\Delta\langle r\rangle > 0$), consistent with DESI including relatively more high-$z$ galaxies than KiDS in those regions.

The specific target selection strategies of DESI also contribute to redshift differences. Figure~\ref{fig:desi_target_types} shows the distribution of the DESI target types across Joint S$_D$ SOM space, with each color scaled for visibility. Notably, there is a high concentration of Emission Line Galaxies (ELGs) (shown in purple) in the lower half of the SOM map, which corresponds to regions where DESI exhibits lower redshift completeness. ELGs, characterized by their distinct emission lines, can present challenges for redshift determination when these features are either too faint or fall outside the DESI spectral coverage. Understanding the relationship between target type distributions and survey completeness across color space is crucial for interpreting biases in redshift calibrations. While ELGs as a target class can achieve very high redshift success rates (often exceeding 95\%), their concentration in specific regions of color-magnitude space means that incomplete sampling in those areas, particularly around \(1.55 < z < 1.65\), can still introduce significant selection effects for lensing surveys. Future studies will further explore the impact of ELG-driven incompleteness on these $n(z)$ distributions.

 The observed limitations in Joint S$_D$’s redshift success rate at higher redshifts are consistent with the DESI spectral design. DESI is optimized for detecting the [O II] doublet (3726.032, 3728.815~\AA), which serves as a primary feature for redshift identification. However, beyond z $\sim$ 1.6, this doublet shifts out of the DESI spectral range due to the 9700~\AA~filter edge. This restriction likely contributes to the reduced success rate at higher redshifts, exacerbating the discrepancy between Joint S$_D$ and deeper surveys such as Joint S$_K$, which incorporates high-redshift sources from zCOSMOS-D. While differences in mean magnitude within a SOM cell may contribute to small mean-redshift shifts (see, e.g., \citealt{Masters2019_ApJ_877_81}; \citealt{McCullough2024}), quantifying this effect requires a cell-dependent estimate of dz/dm. In addition, future work could explore the extent to which the zCOSMOS selection criteria and success rate influence the results.

\section{Summary and Conclusions}
\label{sec:summary_and_conclusions}

In this paper, we used data from DESI DR1 to study the potential of DESI as a robust calibration resource for future lensing surveys. We summarize our key findings below:

\begin{itemize}
    \item The DESI catalog provides over 256k high-quality spectroscopic redshifts (spec-zs) in the COSMOS and XMM/VVDS fields combined \citep{Ratajczak2025}, surpassing previous calibration datasets in size and coverage. DESI re-targets galaxies meaning that the DESI redshifts are fully independent of other previous spectroscopic compilations. Within the KiDS COSMOS footprint, DESI contains approximately 1.6 times more spec-zs than the KiDS spec-z catalog, significantly enhancing the calibration sample in shared regions.
    \item We restrict the analysis to SOM cells jointly populated by both DESI and KiDS and examine how the larger DESI dataset modifies $n(z)$ distributions across five tomographic bins, highlighting its potential as a critical calibration resource for future weak lensing surveys.
    \item Our analysis underscores the importance of spectroscopic completeness in shaping $n(z)$ distributions. 
    As the completeness level increases (e.g., from 60\% to 100\%) per cell of color-space, we observe shifts in the mean redshift on the order of $\Delta z \sim 0.02$. These shifts arise because higher completeness removes galaxies in cells with insufficient spectroscopic sampling, thereby yielding a more accurate representation of the underlying redshift distribution. This highlights the necessity of complete and representative calibration datasets to minimize systematic biases in weak lensing studies.
    \item We find notable differences in the high-redshift tails within the Joint SOM cells, largely driven by the zCOSMOS-Deep sample. One possible explanation is the DESI spectral detection limit of z~$<$~1.6. However, it would also be valuable to investigate the impact of the zCOSMOS-D selection function and redshift success rate. Understanding these high-z tail discrepancies will be crucial for effectively using DESI in photometric redshift calibration as weak lensing is most sensitive to redshift shifts in the highest tomographic bin, where distant galaxies probe more large-scale structure and have the greatest influence on key cosmological parameters like S$_8$.
\end{itemize}

While the goal of this paper has been to showcase the potential of DESI with KiDS-1000, a companion paper~\citep{lange2025} uses these DESI data to recalibrate $n(z)$s for lensing data more generally. In addition, our team is currently working to enhance the consistency of redshift quality criteria used in the DESI calibration sample. In this work, we have adopted the selections of \citet{Ratajczak2025} ensuring that only the most reliable redshift measurements contribute to the calibration process. Importantly, the availability of spectra that do not meet these criteria in the catalog also allows us to assess the effects of redshift selection, an essential step in understanding biases and incompleteness in the calibration sample. Looking forward, we intend to refine our selection process by integrating machine learning (ML) and active learning techniques. These methods will identify high-quality redshifts based on existing VI efforts (Ravulapalli et al in prep). An ultimate goal of this work is to establish a homogeneous catalog with well-defined flags that can be utilized for photo-z calibration. This work is not only instrumental for the DESI collaboration but also sets a precedent for future spectroscopic surveys aiming to achieve high-precision cosmology. These efforts will also ultimately provide a calibration sample for lensing with the Vera C. Rubin Observatory Legacy Survey of Space and Time (LSST).

This work is part of a set of coordinated papers that make up the DESI DR1 Lensing suite. The DR1 analyses include the 3×2-pt cosmology results (Porredon et al., in prep.), the shear+RSD configuration-space cosmology analysis (Semenaite et al., in prep.), the full-scale galaxy–galaxy lensing + clustering cosmology paper (Lange et al., in prep.), and the clustering-redshift validation study (Ruggeri et al., in prep.). Each of these analyses depends on accurate and unbiased redshift-distribution estimates for the DESI weak-lensing source sample. The photometric-redshift calibration developed in this work will contribute to the n(z) inputs for the future DESI Lensing program, which is instrumental in enabling robust cosmological inference across forthcoming analyses.

\section*{ACKNOWLEDGMENTS}
D. Blanco acknowledges support from Cal-Bridge and the U.S. National Science Foundation Graduate Research Fellowship Program (NSF GRFP). We thank the anonymous referee for useful comments on this manuscript.

This material is based upon work supported by the U.S. Department of Energy (DOE), Office of Science, Office of High-Energy Physics, under Contract No. DE–AC02–05CH11231, and by the National Energy Research Scientific Computing Center, a DOE Office of Science User Facility under the same contract. Additional support for DESI was provided by the U.S. National Science Foundation (NSF), Division of Astronomical Sciences under Contract No. AST-0950945 to the NSF’s National Optical-Infrared Astronomy Research Laboratory; the Science and Technology Facilities Council of the United Kingdom; the Gordon and Betty Moore Foundation; the Heising-Simons Foundation; the French Alternative Energies and Atomic Energy Commission (CEA); the National Council of Humanities, Science and Technology of Mexico (CONAHCYT); the Ministry of Science, Innovation and Universities of Spain (MICIU/AEI/10.13039/501100011033), and by the DESI Member Institutions: \url{https://www.desi.lbl.gov/collaborating-institutions}. Any opinions, findings, and conclusions or recommendations expressed in this material are those of the author(s) and do not necessarily reflect the views of the U. S. National Science Foundation, the U. S. Department of Energy, or any of the listed funding agencies.

This research made use of the \texttt{lux} supercomputer at UC Santa Cruz, which is funded by NSF MRI grant AST 1828315.

The authors are honored to be permitted to conduct scientific research on Iolkam Du’ag (Kitt Peak), a mountain with particular significance to the Tohono O’odham Nation.

\section{Data Availability} 
Data points for the figures are available at
https://doi.org/10.5281/zenodo.17959061. 

\appendix

\section{Catalogs used in this paper}\label{app:A}

In this appendix, we present the details of the amalgamation of data discussed in Section ~\ref{sec:data}.

\subsection{Wide Sample}
These are the catalogs used to generate the wide SOM. 
\begin{center}\textit{KiDS-1000 Gold Lensing Catalog}
\end{center} 
Description: The fourth data release of KiDS (KiDS-DR4) comprised of \textit{ugri} imaging and nine-band optical-IR photometry and covering an expanse of 1,000 square degrees. This dataset is characterized by 5-$\sigma$ AB magnitude limits that vary from approximately 25 in the \textit{g} and \textit{r} bands to 23 in \textit{J} and 22 in \textit{K$_s$}. It encompasses a collection of 21,262,011 sources up to redshift $\sim$1, each with reliable shape and redshift measurements, distributed across a grid of 1,006 individual tiles. The COSMOS region data is not integrated into KiDS-DR4. 
\\
References: \cite{Kuijken2019_AA_625_2}, \cite{Wright2020_AA_637_100}, \cite{Giblin2021_AA_645_105}, \cite{Hildebrandt2021_AA_647_124}. \\
Quality Cuts: \texttt{N/A} \\
File name:~\texttt{KiDS\_DR4.1\_ugriZYJHKs\_SOM\_gold\_WL\_cat}
\\~\texttt{.fits}; available on the \href{https://kids.strw.leidenuniv.nl/DR4/KiDS-1000_shearcatalogue.php}{KiDS-1000 data release site}. \\
Nickname: \textbf{W0}

\subsection{Deep Samples}

\begin{center}\textit{COSMOS Deep Sample}
\end{center} 
Description: KiDS + CFHT-z deep photometric catalog in the COSMOS field.
Photo-zs (referred to as z$_B$) are used to establish the tomographic bins utilized in the calibration process. 
It does not include shape information. \\
References: Courtesy of Hendrik Hildebrant and Daniel Gruen. \\
Quality Cuts: \texttt{GAAP\_Flag\_ugriZYJHKs} $== 0$ \\
File name:~\texttt{COSMOSadaptdepth\_ugriZYJHKs\_rot\_photoz}
\\~\texttt{.cat} \\
Nickname: \textbf{D0}

\subsection{Recreating the KiDS Wide SOM}
These are the catalogs used to recreate the KiDS Gold published wide SOM. \\

\begin{center}\textit{KiDS-1000 Gold Catalog SOM Cell Assignments}
\end{center} 
Description: KiDS-1000 gold catalog SOM cell assignments. The KiDS-DR4 SOM cell assignments for each source. This dataset also contains group information, which is the outcome of a hierarchical clustering process applied to the SOM cells, resulting in the creation of approximately 3,000 unique data groupings. \\
Quality Cuts: \texttt{N/A} \\
File name:~\texttt{KiDS\_DR4.1\_ugriZYJHKs\_SOM\_gold\_WL\_cat\ }
\texttt{\_SOMCell.fits} \\
Nickname: \textbf{S0}

\begin{center}\textit{KiDS-1000 SOM Color Vectors}
\end{center} 
Description: Consists of the matrix of color vectors representing the KiDS-1000 Gold SOM cells, which collectively define the position of a cell within an n-dimensional space. \\
Quality flags: \texttt{N/A} \\
File name:~\texttt{K1000\_Spec\_Train\_Adapt\_SOM\_codebook.csv} \\
Nickname: \textbf{S1}

\begin{center}\textit{KiDS-1000 SOM Whitening Parameters}
\end{center} 
Description: The KiDS-1000 Gold SOM codebook vectors undergo training in a whitened color space. This file includes the original whitening parameters. \\
Quality flags: \texttt{N/A} \\
File name:~\texttt{ K1000\_Spec\_Train\_Adapt\_SOM\_whitenparam}
\texttt{.csv} \\
Nickname: \textbf{S2} \\

\bibliographystyle{mnras} 
\bibliography{bibliography}  

\end{document}